\documentclass[twocolumn]{aastex62}

\usepackage{graphicx}	
\usepackage{amsmath}	
\usepackage{amssymb}	
\usepackage[para,flushleft]{threeparttable} 
\usepackage{epstopdf}
\usepackage{textgreek}
\usepackage{comment}
\usepackage{multirow}

\usepackage{xcolor, fontawesome}
\definecolor{twitterblue}{RGB}{64,153,255}
\newcommand{\twitter}[1]{\href{https://twitter.com/#1}{\textcolor{twitterblue}{\faTwitter}\,\tt\hspace{2pt}\textcolor{blue!60!black}{@#1}}}

\hypersetup{colorlinks,linkcolor={red!60!black},citecolor={blue!60!black},urlcolor={blue!60!black}}

\usepackage{natbib,twoopt}
\bibpunct{(}{)}{;}{a}{}{,} 
\makeatletter
\newcommandtwoopt{\citeads}[3][][]{\href{http://ui.adsabs.harvard.edu/\#abs/#3}%
{\def\hyper@linkstart##1##2{}%
\let\hyper@linkend\@empty\citealp[#1][#2]{#3}}}
\newcommandtwoopt{\citepads}[3][][]{\href{http://ui.adsabs.harvard.edu/\#abs/#3}%
{\def\hyper@linkstart##1##2{}%
\let\hyper@linkend\@empty\citep[#1][#2]{#3}}}
\newcommandtwoopt{\citetads}[3][][]{\href{http://ui.adsabs.harvard.edu/\#abs/#3}%
{\def\hyper@linkstart##1##2{}%
\let\hyper@linkend\@empty\citet[#1][#2]{#3}}}
\newcommandtwoopt{\citeyearads}[3][][]%
{\href{http://ui.adsabs.harvard.edu/\#abs/#3}
{\def\hyper@linkstart##1##2{}%
\let\hyper@linkend\@empty\citeyear[#1][#2]{#3}}}
\makeatother

\received{5 March 2020}
\revised{9 April 2020}
\accepted{10 April 2020}
\submitjournal{ApJ}

\shorttitle{Tuning the Exo-Space Weather Radio for Stellar Coronal Mass Ejections}
\shortauthors{Alvarado-G\'omez et al.}

\begin{document}

\title{\sc{Tuning the Exo-Space Weather Radio for Stellar Coronal Mass Ejections}}

\correspondingauthor{Juli\'an D. Alvarado-G\'omez\\ 
\twitter{AstroRaikoh}}
\email{julian.alvarado-gomez@aip.de}

\author[0000-0001-5052-3473]{Juli\'an~D.~Alvarado-G\'omez}

\altaffiliation{Karl Schwarzschild Fellow}
\affil{Leibniz Institute for Astrophysics Potsdam, An der Sternwarte 16, 14482 Potsdam, Germany}

\affil{Center for Astrophysics $|$ Harvard \& Smithsonian, 60 Garden Street, Cambridge, MA 02138, USA}

\author[0000-0002-0210-2276]{Jeremy~J.~Drake}
\affil{Center for Astrophysics $|$ Harvard \& Smithsonian, 60 Garden Street, Cambridge, MA 02138, USA}

\author[0000-0002-5456-4771]{Federico Fraschetti}
\affil{Center for Astrophysics $|$ Harvard \& Smithsonian, 60 Garden Street, Cambridge, MA 02138, USA}
\affil{Dept. of Planetary Sciences-Lunar and Planetary Laboratory, University of Arizona, Tucson, AZ, 85721, USA}

\author[0000-0002-8791-6286]{Cecilia Garraffo}
\affiliation{Institute for Applied Computational Science, Harvard University, Cambridge, MA 02138, USA}
\affil{Center for Astrophysics $|$ Harvard \& Smithsonian, 60 Garden Street, Cambridge, MA 02138, USA}

\author[0000-0003-3721-0215]{Ofer~Cohen}
\affil{University of Massachusetts at Lowell, Department of Physics \& Applied Physics, 600 Suffolk Street, Lowell, MA 01854, USA}

\author[0000-0001-8583-8619]{Christian Vocks}
\affil{Leibniz Institute for Astrophysics Potsdam, An der Sternwarte 16, 14482 Potsdam, Germany}

\author[0000-0003-1231-2194]{Katja Poppenh\"{a}ger}
\affil{Leibniz Institute for Astrophysics Potsdam, An der Sternwarte 16, 14482 Potsdam, Germany}

\author[0000-0002-2470-2109]{Sofia~P.~Moschou}
\affil{Center for Astrophysics $|$ Harvard \& Smithsonian, 60 Garden Street, Cambridge, MA 02138, USA}

\author[0000-0002-9569-2438]{Rakesh K. Yadav}
\affiliation{Department of Earth and Planetary Sciences, Harvard University, Cambridge, MA 02138, USA}

\author[0000-0003-0472-9408]{Ward B. Manchester IV}
\affiliation{Department of Climate and Space Sciences and Engineering, University of Michigan, Ann Arbor, MI 48109, USA}



\begin{abstract}

\noindent Coronal mass ejections (CMEs) on stars other than the Sun have proven very difficult to detect. One promising pathway lies in the detection of type~II radio bursts. Their appearance and distinctive properties are associated with the development of an outward propagating CME-driven shock. However, dedicated radio searches have not been able to identify these transient features in other stars. Large Alfv\'en speeds and the magnetic suppression of CMEs in active stars have been proposed to render stellar eruptions ``radio-quiet''. Employing 3D magnetohydrodynamic simulations, we study here the distribution of the coronal Alfv\'en speed, focusing on two cases representative of a young Sun-like star and a mid-activity M-dwarf (Proxima Centauri). These results are compared with a standard solar simulation and used to characterize the shock-prone regions in the stellar corona and wind. Furthermore, using a flux-rope eruption model, we drive realistic CME events within our M-dwarf simulation. We consider eruptions with different energies to probe the regimes of weak and partial CME magnetic confinement. While these CMEs are able to generate shocks in the corona, those are pushed much farther out compared to their solar counterparts. This drastically reduces the resulting type II radio burst frequencies down to the ionospheric cutoff, which impedes their detection with ground-based instrumentation. 

\end{abstract}




\section{Introduction}\label{sec:intro}

\noindent Flares and coronal mass ejections (CMEs) are more energetic than any other class of solar phenomena. These events involve the rapid (seconds to hours) release of up to $10^{33}$~erg of magnetic energy in the form of particle acceleration, heating, radiation, and bulk plasma motion (\citeads{2012LRSP....9....3W}, \citeads{2017LRSP...14....2B}). Displaying much larger energies (by several orders of magnitude), their stellar counterparts are expected to play a fundamental role in shaping the evolution of activity and rotation (\citeads{2013ApJ...764..170D}, \citeads{2017ApJ...840..114C}, \citeads{2017MNRAS.472..876O}), as well as the environmental conditions around low-mass stars (see e.g.,~\citeads{2018haex.bookE..19M}). Energetic photon and particle radiation associated with flares and CMEs are also the dominant factors driving evaporation, erosion, and chemistry of protoplanetary disks (e.g.,~\citeads{1997ApJ...480..344G}, \citeads{2009ApJ...703.2152T}, \citeads{2018ApJ...853..112F}) and planetary atmospheres (e.g.,~\citeads{2013oepa.book.....L}, \citeads{2017ApJ...846...31C}, \citeads{2019AsBio..19...64T}). This is critical for exoplanets in the close-in habitable zones around M-dwarfs, which are the focus of recent efforts on locating nearby habitable planets (\citeads{2008PASP..120..317N}, \citeads{2019arXiv190604644T}), but are known for their long sustained periods of high flare activity (see \citeads{2016hasa.book...23O}, \citeads{2019ApJ...871..241D}).  

Stellar flares are now routinely detected across all  wavelengths from radio to X-ray, spectral types from F-type to brown dwarfs, and ages from stellar birth to old disk populations (e.g., \citeads{2016ApJ...829...23D}; \citeads{2019A&A...622A.133I}; \citeads{2019A&A...622A.210G}). This wealth of information is increasingly driving the study of their effects on exoplanet atmospheres (e.g., \citeads{2010AsBio..10..751S}, \citeads{2018ApJ...865..101M}, \citeads{2019AsBio..19...64T}).

On the other hand, the observational evidence for stellar CMEs is very thin, with a single direct detection of an extreme event recently reported by \citetads{2019NatAs.tmp..328A} using \textit{Chandra}. Unfortunately, current X-ray instrumentation renders the methodology of this detection\,---resolving (temporally and spectroscopically) a post-flare blueshift signature in cool coronal lines---\,sensitive only to the most energetic eruptions. As described by \citetads{2019ApJ...877..105M}, other diagnostics, such as asymmetries in Balmer lines or continued X-ray absorption during flares, have provided only a handful of good CME candidates so far (see also \citeads{2017ApJ...850..191M}, \citeads{2019A&A...623A..49V}).  

In analogy with the Sun, an alternative way of recognizing CMEs in distant stars lies in the detection of the so-called type II radio bursts (\citeads{1950AuSRA...3..387W}, \citeads{1965sra..book.....K}, \citeads{2017IAUS..328..243O}). These transients correspond to two distinct bright lanes in radio spectra in the kHz - MHz range, separated by a factor of $\sim2$ in frequency, characterized by a gradual drift from high to low frequencies. These features are attributed to emission at the fundamental and first harmonic of the plasma frequency\footnote[6]{Expressed in c.g.s. units as $\nu_{\rm p} = (2\pi)^{-1}(\sqrt{4\pi e^2/m_{\rm e}})\sqrt{n}$, where $e$, $m$ are the electron charge and mass, and $n$ denotes the number density of the ambient region.}, resulting from non-thermal electrons accelerated by a shock generated as the velocity of the CME in the stellar wind frame surpasses the local Alfv\'en speed (i.e., $U^{\rm CME} - V^{\rm SW} > V_{\rm A} = B/\sqrt{4\pi \rho}$, with $V^{\rm SW}$, $B$ and $\rho$ as the stellar wind speed, magnetic field strength and plasma density, respectively\footnote[7]{In a low density/strong field regime, the expression for the Alfv\'en speed should be modified as $V_{\rm A} = c /\sqrt{1 + 4\pi\rho c^{2}/B^{2}}$, where $c$ is the speed of light. This prevents $V_{\rm A}$ to be larger than~$c$.}). The frequency drift reflects the decrease in particle density with distance as the CME shock propagates outward in the corona (see \citeads{2003SSRv..107...27C}, \citeads{2006cme..book..341P} and references therein).  

While solar observations indicate an association rate of just $\sim$\,$1-4$\,\% between CMEs and type II bursts in general (see \citeads{2005JGRA..11012S07G}, \citeads{2018Ge&Ae..58..989B}), it is close to $100$\,\% for the most energetic eruptions (Gopalswamy et~al. \citeyearads{2009SoPh..259..227G}, \citeyearads{2019arXiv191207370G}). Furthermore, the high fraction of CMEs associated with strong flares in the Sun ($\sim$$80-90$\,\% for X-class flares, see \citeads{2009IAUS..257..233Y}, \citeads{2017SoPh..292....5C}), combined with the enhanced stellar flare rates and energies (e.g., \citeads{2002ApJ...580.1118K}, \citeads{2007A&A...471..645C}, \citeads{2013ApJS..209....5S}, \citeads{2018ApJ...867...71L}), are expected to yield enough type II burst events in active stars to secure detections. While several other radio transients have been detected in low-mass stars (see e.g.,~\citeads{2019ApJ...871..214V}), this particular class of radio events has not been observed so far (e.g.,~\citeads{2016ApJ...830...24C}, \citeads{2017PhDT.........8V}, Crosley \& Osten \citeyearads{2018ApJ...856...39C}, \citeyearads{2018ApJ...862..113C}). However, based on solar statistics it is clear that the lack of type~II bursts does not imply an absence of CMEs. 

Recent numerical studies are starting to provide a common framework to interpret these observations and the apparent imbalance between flare and CME occurrence in stars. Detailed MHD models have shown that the stellar large-scale magnetic field can establish a suppression threshold preventing CMEs of certain energies from escaping (\citeads{2016IAUS..320..196D}, \citeads{2018ApJ...862...93A}). The rationale of this mechanism comes from solar observations, where it has been proposed to operate on smaller scales, forming a magnetic cage for the plasma ejecta in certain flare-rich CME-poor active regions (e.g.~Thalmann~et~al.~\citeyearads{2015ApJ...801L..23T}, \citeyearads{2017ApJ...844L..27T}, \citeads{2015ApJ...804L..28S}), preventing access to open-field sectors that could facilitate breakout (\citeads{2016ApJ...826..119L}, \citeads{2018ApJ...861..131D}). 

The results of \citetads{2018ApJ...862...93A} also predict that due to magnetic suppression, escaping stellar CMEs would be slower and less energetic compared to similar events occurring under weaker (or negligible) large-scale fields. In a recent observational study performed by \citetads{2019ApJ...871..214V}, low CME velocities compared with the local Alfv\'en speed were considered as a possible explanation for the lack of type~II events in very active, radio bursting M-dwarfs. Using 1D models and scaling laws, \citetads{2019ApJ...873....1M} suggested that CMEs in M-dwarfs would be radio-quiet, as they would not be able to overcome the large Alfv\'en speeds in the corona that are the product of the strong surface magnetic fields present on these stars (see \citeads{2011IAUS..271...23D}, \citeads{2014IAUS..302..156R}, \citeads{2019A&A...626A..86S}).

Here, we expand and complement these previous ideas, analyzing the Alfv\'en speed distributions resulting from realistic 3D state-of-the-art corona and stellar wind models. We compare results for the Sun during an active period, for a large-scale dipole-dominated geometry representative of a young Sun-like star, and for a high-complexity strong field distribution predicted from a dynamo simulation of a fully convective M-dwarf (Proxima Centauri, \citeads{2016ApJ...833L..28Y}). Furthermore, we test the hypothesis of radio-quiet CMEs in M-dwarfs by simulating eruptions in the regimes of weak and partial large-scale magnetic field confinement, and determine whether or not they become super-Alfv\'enic during their temporal evolution. 

This paper is organized as follows: Section \ref{sec:models} contains a description of the employed models and boundary conditions. Results from the steady-state Alfv\'en speed distributions, as well as the time-dependent M-dwarf CME simulations, are presented in Section~\ref{sec:results}. We discuss our findings and their implications in Section~\ref{sec:discussion}, and provide a brief summary in Section~\ref{sec:summary}. 

\section{Models}\label{sec:models}

\noindent The numerical simulations discussed in this work follow closely the methodology described in Alvarado-G\'omez~et~al. (\citeyearads{2018ApJ...862...93A}, \citeyearads{2019ApJ...884L..13A}), where different models included in the \href{http://csem.engin.umich.edu/tools/swmf/}{Space Weather Modeling Framework} (SWMF, \citeads{2018LRSP...15....4G}) are used. 

\subsection{Steady-State Configurations}

\noindent The corona and stellar wind solutions are based on the 3D MHD code BATS-R-US \citepads{1999JCoPh.154..284P,2012JCoPh.231..870T} and the data-driven Alfv\'en Wave Solar Model (AWSoM, \citeads{2014ApJ...782...81V}). The latter, extensively validated and updated against remote and in-situ solar data (e.g.,~\citeads{2017ApJ...845...98O}, \citeads{2019ApJ...872L..18V}, \citeads{2019ApJ...887...83S}), considers Alfv\'en wave turbulent dissipation as the main driver of coronal heating and stellar wind acceleration. Both contributions are self-consistently calculated and incorporated as additional source terms in the energy and momentum equations, which, combined with the mass conservation and magnetic field induction equations, close the non-ideal MHD set of equations solved numerically. Radiative losses and effects from electron heat conduction are also taken into account. Our simulation domain extends from $\sim$\,$1\,R_{\bigstar}$ to $85\,R_{\bigstar}$, and employs a radially-stretched spherical grid with a maximum base resolution of $0.025\,R_{\bigstar}$, with the stellar rotation axis aligned with the $z$ cartesian direction. 

The simulation evolves until a steady-state is reached. For solar models, the distribution of the photospheric magnetic field averaged over one rotation (known as a synoptic magnetogram), serves as the inner boundary condition from which the Alfv\'en wave dissipation spectrum is constructed (see \citeads{2014ApJ...782...81V} for details). Our solar run is based on the synoptic magnetogram\footnote[8]{Acquired by the \href{https://gong.nso.edu/data/magmap/}{Global Oscillation Network Group (GONG).}} associated with Carrington rotation (CR) 2107 (Feb./Mar. 2011, rising phase of cycle 24). We use this particular CR as its resulting AWSoM solution has been well studied in previous numerical works (e.g., \citeads{2013ApJ...764...23S}, Jin~et~al.~\citeyearads{2017ApJ...834..172J}, \citeyearads{2017ApJ...834..173J}). However, for the purposes of this study, any CR with the presence of AR groups could serve to drive the reference model. Apart from the CR magnetogram, solar chromospheric levels of plasma density ($n_{0} = 2\times10^{10}$~cm$^{-3}$) and temperature ($T_{0} = 5\times10^{4}$~K) are also set at the simulation inner boundary. Default values are used for the remaining parameters of AWSoM. This includes the proportionality constants for the Alfv\'en wave Poynting flux $(S/B)_{\bigstar} = 1.1\times10^{6}$~W~m$^{-2}$~T$^{-1}$, and correlation length $L_{\perp}\sqrt{B} = 1.5\times10^{5}$~m~$\sqrt{\text{T}}$ (Table~1 in \citeads{2014ApJ...782...81V}; see also \citeads{2013ApJ...764...23S} for more details). 

As mentioned earlier, these boundary conditions have been thoroughly tested in several AWSoM validations. We preserve all of them in our Sun-like dipole dominated case, modifying only the surface field distribution to include a large-scale 75~G dipolar component (as described in \citeads{2018ApJ...862...93A}). For this stellar model (and our reference simulation of the Sun), we assumed fiducial solar values of mass ($M_{\bigstar} = M_{\odot}$), radius ($R_{\bigstar} = R_{\odot}$), and rotation period ($P_{\rm rot} = 25.38$~d). For the comparative analysis presented here, we do not consider the influence of a shorter rotation period expected from a younger Sun. Still, as noted in \citetads{2018ApJ...862...93A}, the resulting AWSoM solution in this case is consistent with observational constraints of stellar winds in young late-type stars (see \citeads{2005ApJ...628L.143W}), which display comparable field strengths and geometries to the one assumed here (e.g. $\xi$ Boo A, \citeads{2012A&A...540A.138M}; $\epsilon$ Eri, Jeffers~et~al.~\citeyearads{2014A&A...569A..79J}, \citeyearads{2017MNRAS.471L..96J}).

The boundary conditions for the M-dwarf regime are much less constrained by observations. Here we employ the same driving conditions as in \citetads{2019ApJ...884L..13A}, using the field topology emerging at the cyclic regime of a 3D dynamo simulation tailored to Proxima Centauri (\citeads{2016ApJ...833L..28Y}). We scale the surface radial field between $\pm1400$~G, which yields an average field strength compatible with the lower bound from Zeeman broadening measurements on this star ($600 \pm 150$~G, \citeads{2008A&A...489L..45R}) of $\sim$\,$450$~G. 

The presence of strong and complex surface magnetic fields is expected to drastically influence the coronal structure and stellar wind  (e.g.~\citeads{2014MNRAS.438.1162V}, \citeads{2016A&A...595A.110G}, \citeads{2017ApJ...834...14C}). 
\newpage

\setcounter{page}{5}
\begin{figure}[!h]
\centering
\includegraphics[trim=0.0cm 0.0cm 0.0cm -8.85cm, clip=true,width=0.382\textwidth]{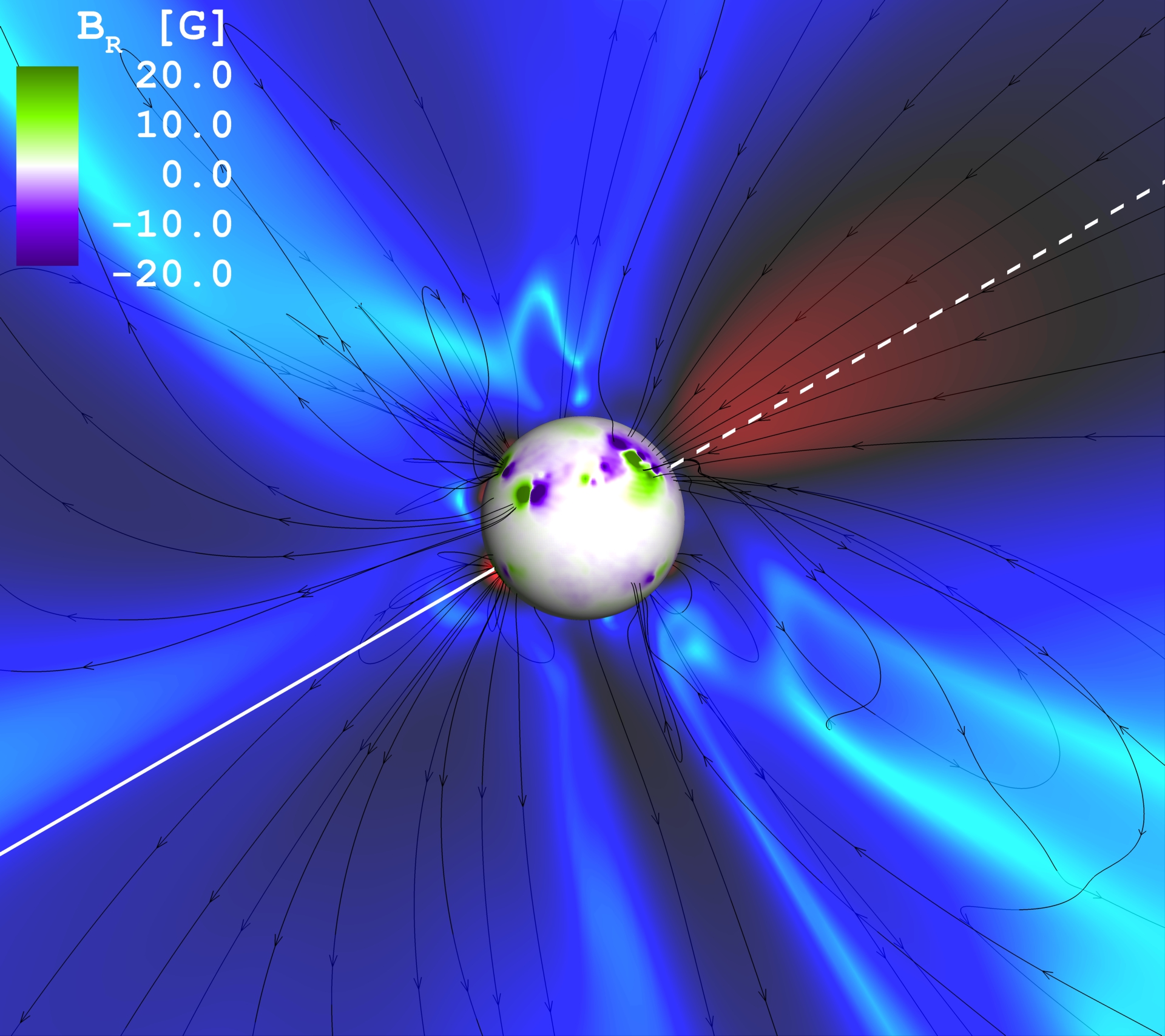}\vspace{1pt}
\includegraphics[trim=0.0cm 0.0cm 0.0cm 0.0cm, clip=true,width=0.382\textwidth]{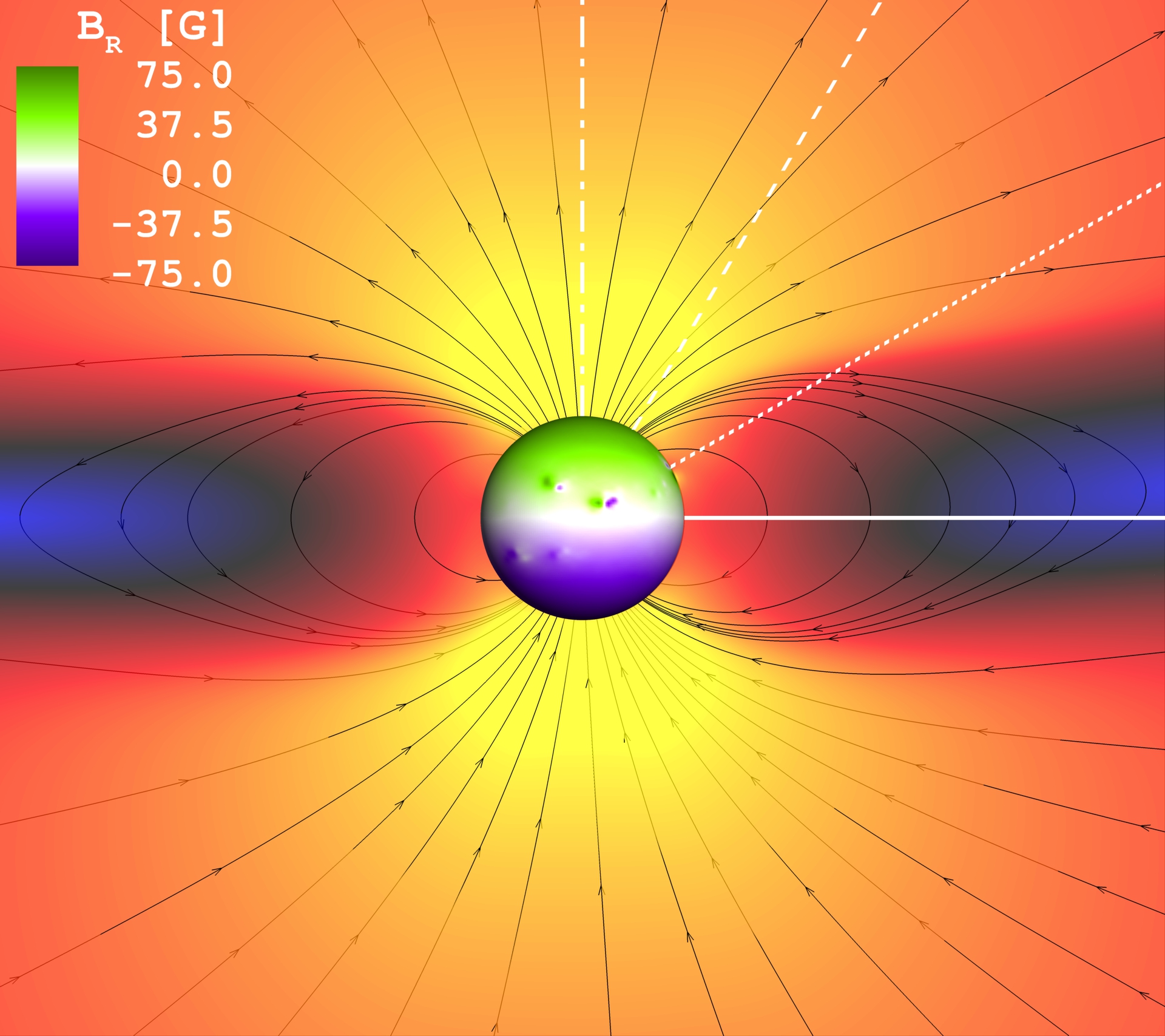}\vspace{1pt}
\includegraphics[trim=0.0cm 0.0cm 0.0cm 0.0cm, clip=true,width=0.382\textwidth]{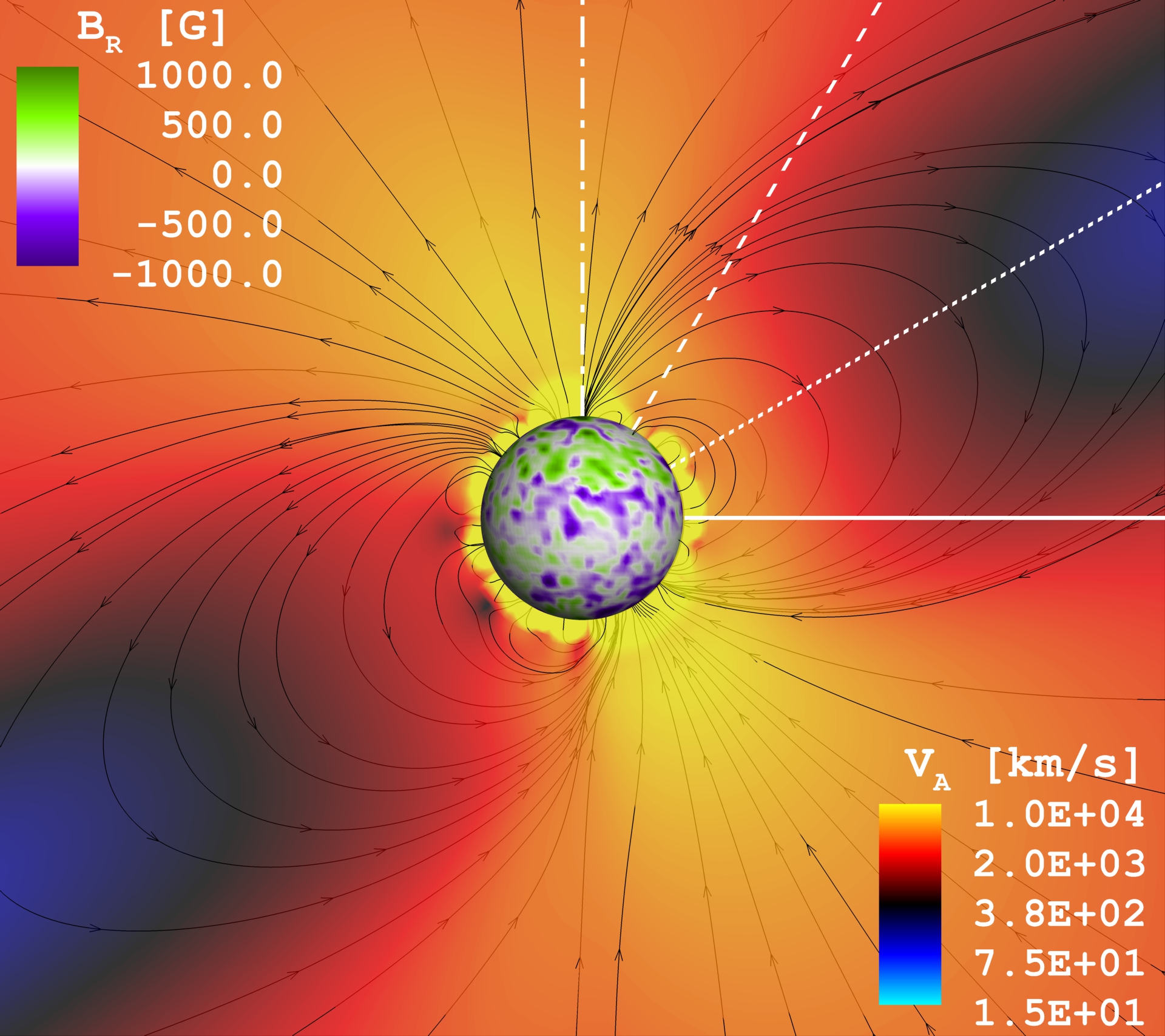}
\caption{Meridional projection of the Alfv\'en speed ($V_{\rm A}$) in our AWSoM steady-state simulations. \textit{Top}: Sun (CR\,2107); \textit{Middle}: Young Sun-like star (CR\,2107 + 75~G large-scale dipole); \textit{Bottom}: M-dwarf (Proxima Centauri). The sphere represents the stellar surface, color-coded by the radial field ($B_{\rm R}$) driving each model. The color scaling for $V_{\rm A}$ is preserved in all cases. The $V_{\rm A}$ profiles in Fig.~\ref{fig_2} have been extracted along the white lines indicated. The field of view is 12\,$R_{\bigstar}$ with a set of selected magnetic field lines in black.}
\label{fig_1}
\end{figure}

\begin{figure}[!h]
\centering
\includegraphics[trim=1.2cm 2.4cm 0.9cm 0.1cm, clip=true,width=0.485\textwidth]{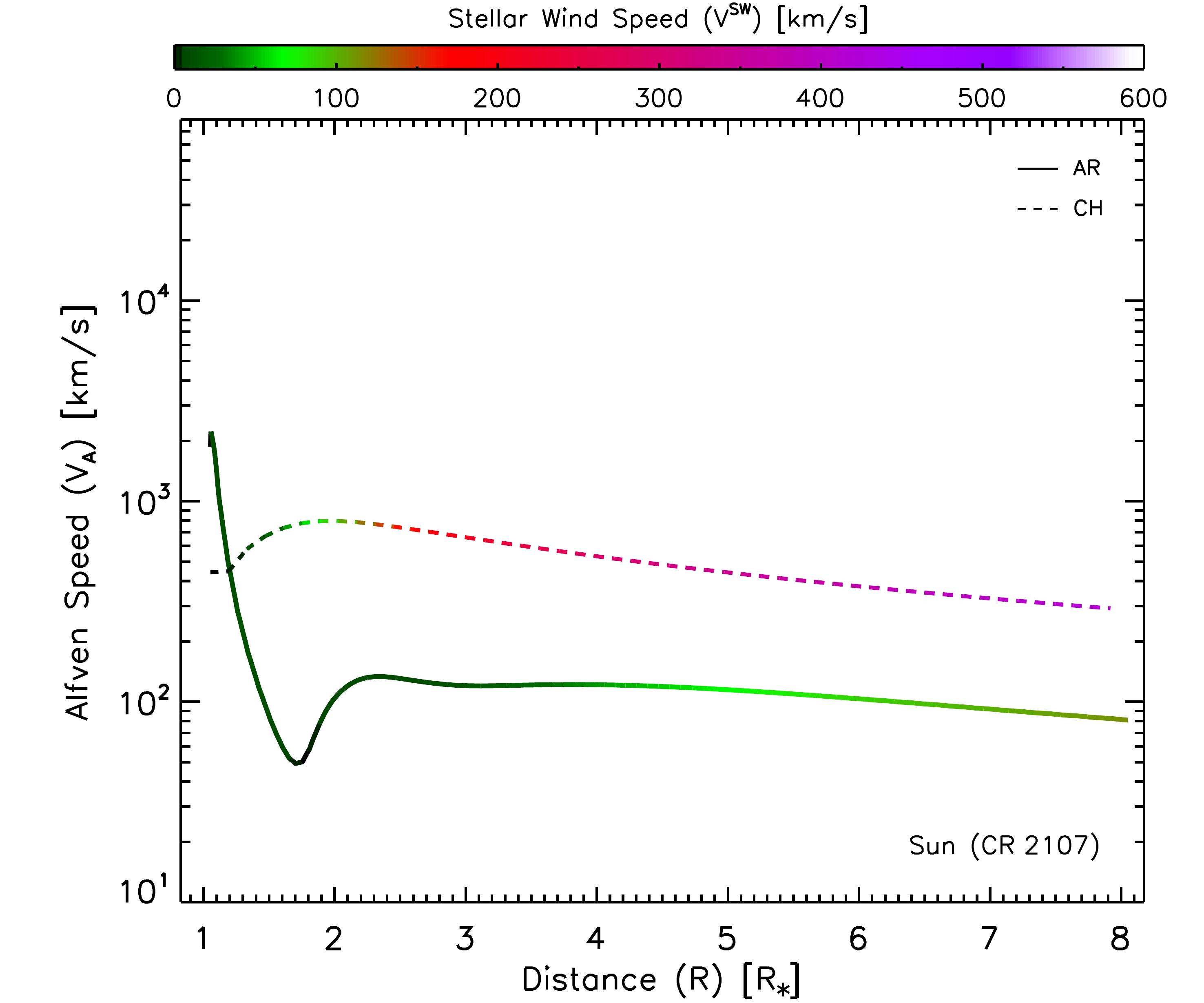}
\includegraphics[trim=1.2cm 2.4cm 0.9cm 2.922cm, clip=true,width=0.485\textwidth]{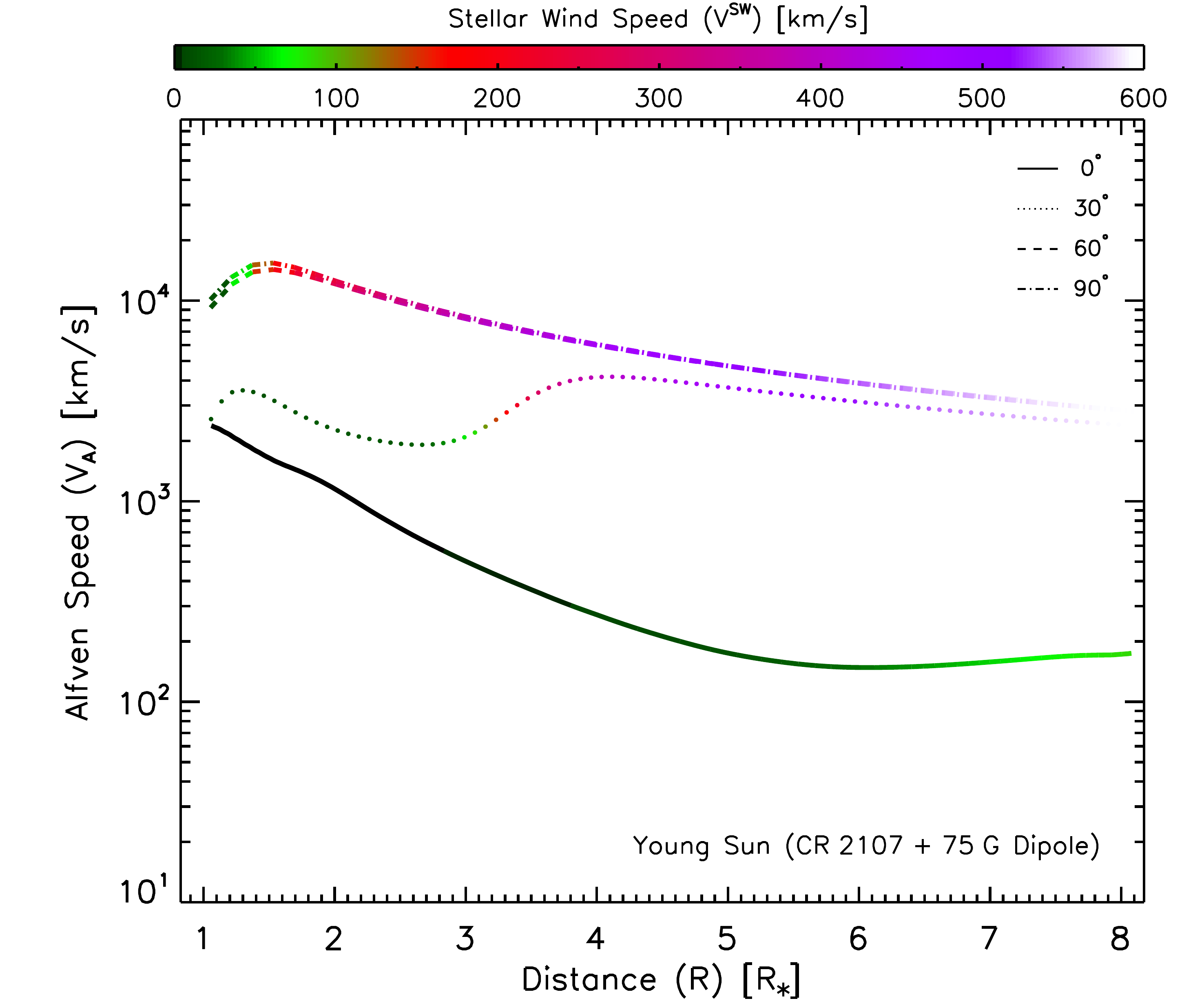}
\includegraphics[trim=1.2cm 0.1cm 0.9cm 2.922cm, clip=true,width=0.485\textwidth]{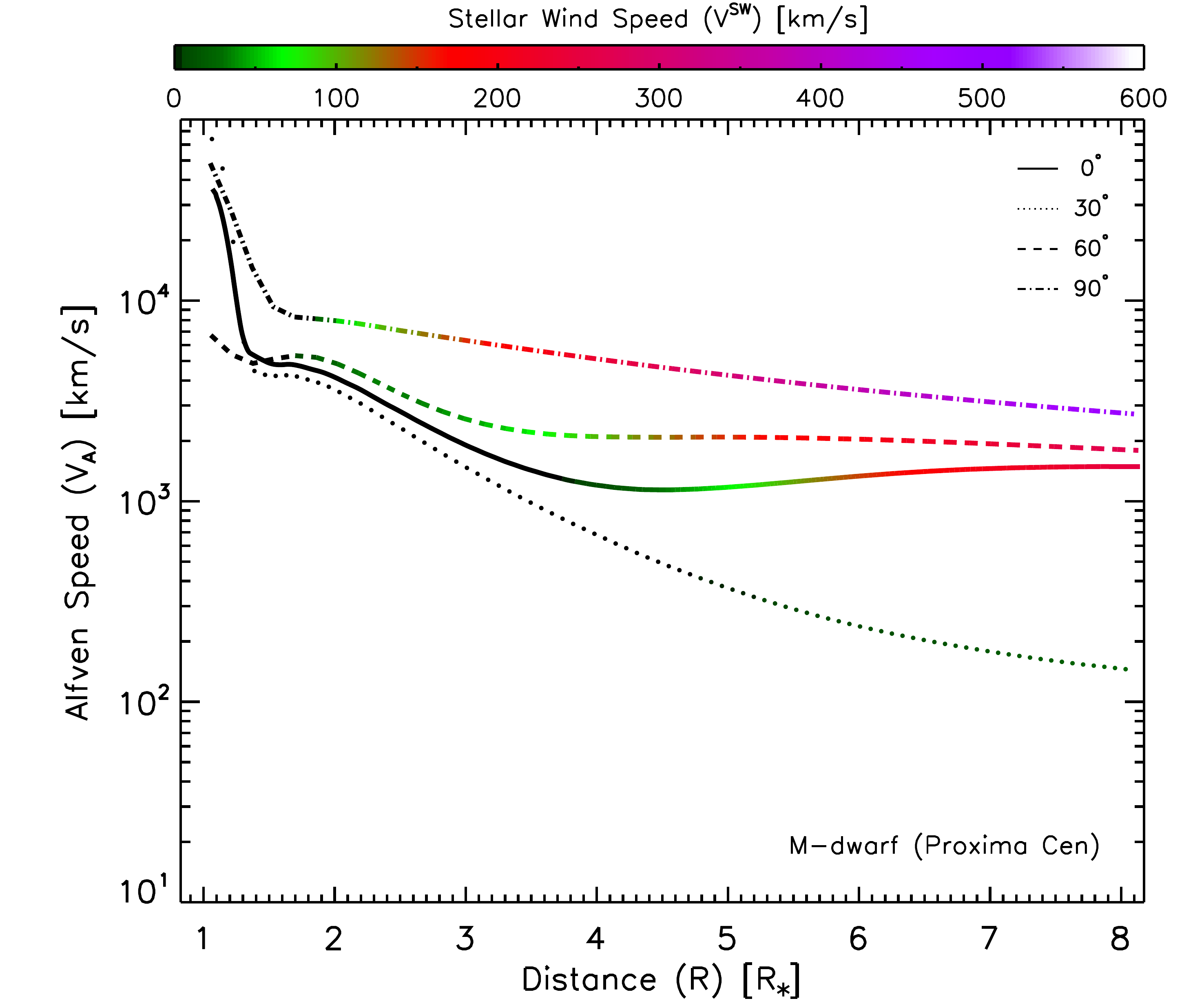}
\caption{Radial profiles of $V_{\rm A}$ extracted from each of the AWSoM steady-state solutions (Fig.~\ref{fig_1}, white lines). Active Region (AR) and Coronal Hole (CH) profiles are included for the solar case (top panel). Four latitudes are probed in our two stellar cases (middle and bottom panels). The stellar wind speed ($V^{\rm SW}$) along each profile is indicated by the color-scale.}
\label{fig_2}
\end{figure}
\clearpage

\begin{deluxetable*}{lCCCCCCCCC}[ht!]
\tablecaption{Set of parameters assumed in the steady-state simulations using AWSoM.\label{tab_1}}
\tablecolumns{10}
\tablenum{1}
\tablewidth{0pt}
\tablehead{
\colhead{Model} & \colhead{Magnetic Field} & \colhead{$\left<B\right>_{\rm S}$} & \colhead{$M_{\bigstar}$} & \colhead{$R_{\bigstar}$} & \colhead{$P_{\rm rot}$} & \colhead{$n_{0}$} & \colhead{$T_{0}$} & \colhead{$(S/B)_{\bigstar}$} & \colhead{$L_{\perp}\sqrt{B}$}\vspace{-0.15cm}\\
\colhead{} & \colhead{Distribution} & \colhead{[G]} & \colhead{[$M_{\odot}$]} & \colhead{[$R_{\odot}$]} & \colhead{[d]} & \colhead{[cm$^{-3}$]} & \colhead{[K]} & \colhead{[W~m$^{-2}$~T$^{-1}$]} & \colhead{[m~$\sqrt{\text{T}}$]}  
}
\startdata
Sun & \text{CR\,2107} & 3.0 & 1.0 & 1.0 & 25.38 & 2\times10^{10} & 5\times10^{4} & 1.1\times10^{6} & 1.5\times10^{5} \\
Sun-like & \text{CR\,2107\,+\,(75~G)}^{\dagger} & 42.6 & 1.0 & 1.0 & 25.38 & 2\times10^{10}$ & 5\times10^{4} & 1.1\times10^{6} & 1.5\times10^{5} \\
M-dwarf & \text{Proxima Centauri}^{\ddagger} & 448.8 & 0.122 & 0.155 & 83.0 & 2\times10^{10}$ & 5\times10^{4} & 1.1\times10^{6} & 6.0\times10^{5} \\
\enddata
\tablenotetext{}{$^{\dagger }$Added to the first term in the spherical harmonic expansion of the surface magnetic field (large-scale dipole).}
\vspace{-0.12cm}\tablenotetext{}{$^{\ddagger }$Snapshot at $490$ rotations from the high-resolution dynamo simulation of \citetads{2016ApJ...833L..28Y}.}
\end{deluxetable*}

From the AWSoM perspective, this could imply that additional modifications (apart from the surface magnetogram) to the boundary conditions might be required. However, as discussed by \citeads{2013ApJ...764...23S}, the scaling $(S/B)_{\bigstar}$ in the Alf\'ven wave Poynting flux is consistent and equivalent to the magnetic to X-ray flux empirical relation of \citetads{2003ApJ...598.1387P}, which extends beyond the magnetic fluxes observed in M-dwarfs stars. For this reason, we retain the standard AWSoM value for this parameter in our Proxima Centauri simulations. 

On the other hand, we consider the currently available information on M-dwarf stellar winds to adjust the normalization for the Alfv\'en wave correlation length $L_{\perp}\sqrt{B}$. Unfortunately, stellar wind properties in low mass stars (particularly M-dwarfs) are highly uncertain. Estimates of mass loss rates ($\dot{M}$)\,---interpreted as a measure of wind strength---\,are only available for 14 stellar systems (12 detections, 2 upper limits), of which only 3 are M-dwarfs (2 detections, 1 upper limit; see \citeads{2018JPhCS1100a2028W}). For the particular case of Proxima Centauri, two different methods have been used to constrain the mass loss rate associated with its stellar wind. An upper limit of $\dot{\rm M} < 0.2~\dot{\rm M}_{\odot}$ was placed by \citetads{2001ApJ...547L..49W}, through the astrospheric signature in the Ly$\alpha$ line (\citeads{2014ASTRP...1...43L}). A higher limit of $\dot{\rm M} < 14$~$\dot{\rm M}_{\odot}$ was found by \citetads{2002ApJ...578..503W}, measuring in X-rays the direct signature of charge exchange between the stellar wind ions and the local interstellar medium.
\setcounter{page}{4}

As the astrospheric method has been more commonly applied, we set\footnote[9]{For completeness, we also performed a simulation using the default AWSoM value for $L_{\perp}\sqrt{B}$. The resulting stellar wind mass loss rate in this case is $\dot{\rm M} \simeq 0.09$~$\dot{\rm M}_{\odot}$.} $L_{\perp}\sqrt{B} = 6.0 \times 10^{5}$~m~$\sqrt{\text{T}}$ which yields a stellar wind mass loss rate of $\dot{\rm M} \simeq 0.3$~$\dot{\rm M}_{\odot}$, which is still consistent with the Ly$\alpha$ limit (taking into account the typical errors of this technique; see \citeads{2014ASTRP...1...43L}). This is the same $L_{\perp}\sqrt{B}$ value used in the stellar wind simulations of Proxima Centauri by \citetads{2016ApJ...833L...4G}, and Barnard's Star by \citetads{2019ApJ...875L..12A}. Finally, published stellar properties for this object are used in this case ($M_{\bigstar} = 0.122$~$M_{\odot}$, $R_{\bigstar} = 0.154$~$R_{\odot}$, $P_{\rm rot} = 83.0$~d, \citeads{2007AcA....57..149K}, \citeads{2017A&A...598L...7K}, \citeads{2017A&A...602A..48C}). Table \ref{tab_1} contains a summary of all the parameters considered in our AWSoM steady-state simulations.

\subsection{Flux-Rope CME Model}\label{sec:FR}

\noindent The \citetads[TD,]{1999A&A...351..707T} flux-rope eruption model is used to drive our M-dwarf CME simulations. In the SWMF implementation (e.g.,~\citeads{2008ApJ...684.1448M}, \citeads{2013ApJ...773...50J}), the twisted loop-like structure of the TD model is coupled to the AWSoM steady-state solution at the inner boundary (stellar surface), and initialized with eight different parameters related to the location (2), orientation, size (3), magnetic free energy ($E_{\rm B}^{\rm FR}$) and mass of the flux-rope ($M^{\rm FR}$). The CME simulations discussed here use the same parameters as in \citetads{2019ApJ...884L..13A}, namely, longitude ($270$~deg), latitude ($36$~deg), tilt angle ($28$~deg), flux-tube radius ($20$~Mm), length ($\sim$\,$150$~Mm), and loaded mass ($M^{\rm FR} = 4.0 \times~10^{14}$~g). We consider two values of $E_{\rm B}^{\rm FR}$ in order to probe the regimes of weak ($E_{\rm B,1}^{\rm FR} \simeq 4.1 \times 10^{35}$~erg) and partial ($E_{\rm B,2}^{\rm FR} \simeq 2.0 \times 10^{35}$~erg) large-scale magnetic CME confinement\footnote[10]{This was necessary as in the Proxima-like $\pm1400$~G surface field scaling employed here, the CME event simulated in \citetads{2019ApJ...884L..13A}, with $E_{\rm B}^{\rm FR}~\simeq 6.57 \times 10^{34}$~erg, was fully confined by the large-scale magnetic field.}. Note that assuming a similar flare-CME magnetic energy partition as in the Sun (e.g., \citeads{2012ApJ...759...71E}, \citeads{2017ApJ...834...56T}), our selected $E_{\rm B}^{\rm FR}$ values are consistent and sufficient to power the best CME candidate observed in Proxima Centauri so far ($F_{\rm X} \simeq 1.7 \times 10^{31}$~erg, $E_{\rm K}^{\rm CME} \simeq 5 \times 10^{31}$~erg; see \citeads{2019ApJ...877..105M}). The initial parameters assumed in our TD flux-rope simulations are listed in Table~\ref{tab_2}.

Each CME simulation evolves for $90$ minutes (real time) from which we extract 3D snapshots of the entire simulation domain at a cadence of $1$ minute.     

\begin{deluxetable}{lcc}[ht!]
\tablecaption{Flux rope parameters initializing the TD CME simulations within the M-dwarf model.\label{tab_2}}
\tablecolumns{3}
\tablenum{2}
\tablewidth{2pt}
\tablehead{\colhead{Parameter} & \colhead{Value} & \colhead{Unit}}
\startdata
Latitude & 36.0 & deg\\
Longitude & 270.0 & deg\\
Tilt angle$^{\dagger}$ & 28.0 & deg\\
Radius ($R^{\rm FR}$) & 20.0 & Mm\\
Length ($L^{\rm FR}$) & 150.0 & Mm \\
Mass ($M^{\rm FR}$) & $4.0\times~10^{14}$ & g\smallskip\\
\hline
\multirow{2}{*}{Magnetic energy ($E_{\rm B}^{\rm FR}$)} & \tablenotemark{a}$4.1 \times 10^{35}$ & \multirow{2}{*}{erg}\\ 
& \tablenotemark{b}$2.0 \times 10^{35}$ &\\
\enddata
\tablenotetext{}{\hspace{-0.2cm}$^{\dagger }$Measured with respect to the stellar equator in the counter clock-wise direction.}
\vspace{-0.04cm}\tablenotetext{}{\hspace{-0.2cm}$^{\rm a }$Weakly confined CME.}
\vspace{-0.12cm}\tablenotetext{}{\hspace{-0.2cm}$^{\rm b }$Partially confined CME.}
\end{deluxetable}

\section{Results}\label{sec:results}

\subsection{Coronal Alfv\'en Speed Profiles}\label{sec:ss}

\noindent We examine first the coronal Alfv\'en speed distributions obtained in our steady-state solutions. Figure~\ref{fig_1} compares the resulting $V_{\rm A}$ on an arbitrary meridional projection, with a common color scale saturated between $15$~km~s$^{-1}$~$\le V_{\rm A} \le 10000$~km~s$^{-1}$. Radial profiles for different regions/latitudes are indicated and correspondingly plotted in Fig.~\ref{fig_2}, together with the stellar wind speed ($V^{\rm SW}$) along the same profiles. 

As expected, our simulated values of $V_{\rm A}$ for the Sun\,---which are consistent with observations (see \citeads{2014A&A...564A..47Z})---\,are globally lower than their stellar counterparts (by up to one order of magnitude). The spatial distribution of $V_{\rm A}$ also follows the nominal behavior in the solar corona: large $V_{\rm A}$ values very close to active regions (strong small-scale field) and above coronal holes (low density sectors; see also Fig.~\ref{fig_2}, top panel). The lower $V_{\rm A}$ locations coincide with the high-density coronal streamers. Furthermore, the appearance of local minima in $V_{\rm A}$ above active regions, due to the superposition of the small- and large-scale solar magnetic field, is also captured in our simulation (see \citeads{2003A&A...400..329M}).

\begin{figure*}[!t]
\centering
\includegraphics[trim=0.0cm 0.1cm 0.1cm 0.0cm, clip=true,width=0.995\textwidth]{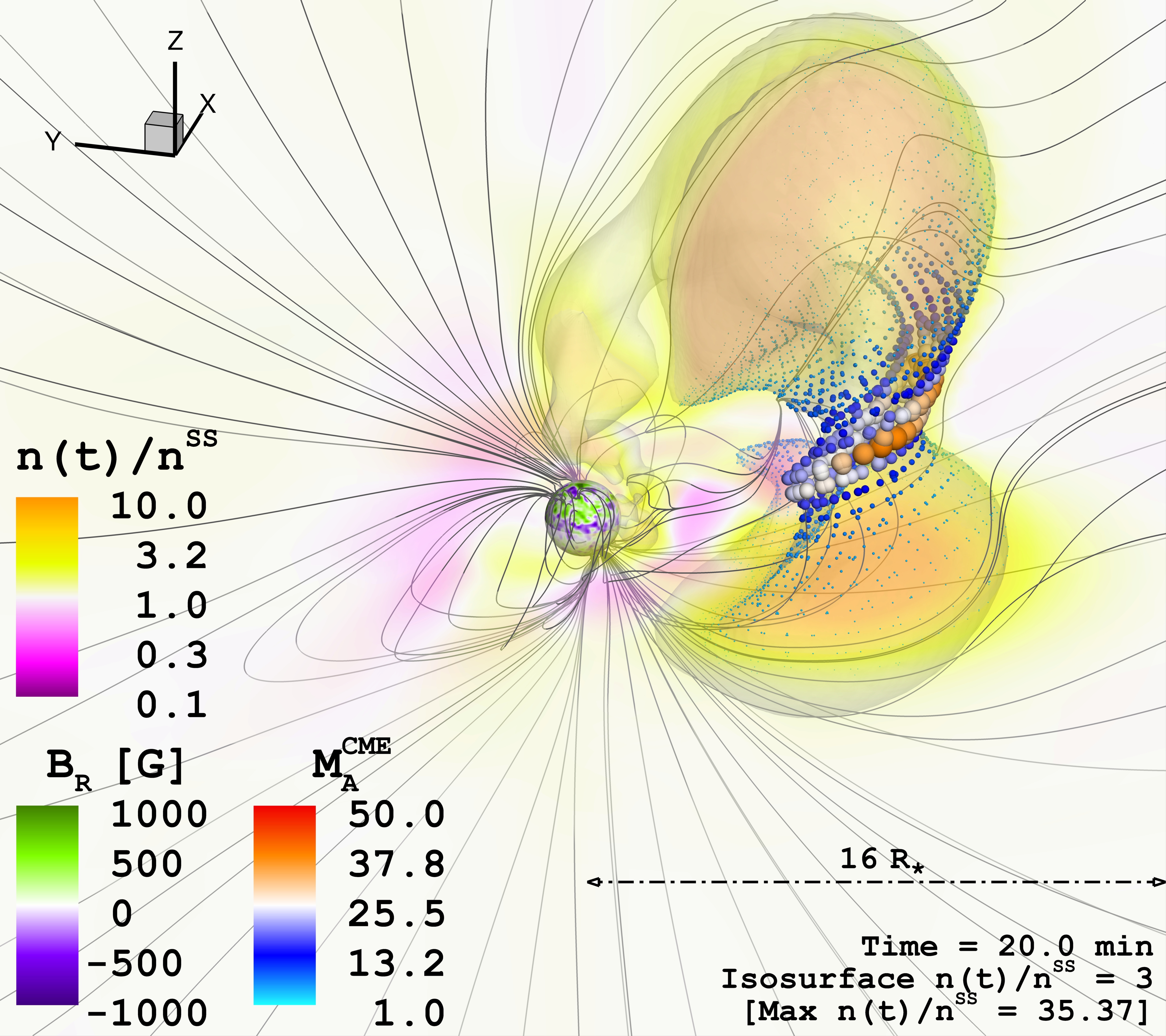}
\caption{Snapshot during the temporal evolution of the weakly confined CME ($E_{\rm B,1}^{\rm FR} \simeq 4.1 \times 10^{35}$~erg, $E_{\rm K,1}^{\rm CME} \simeq 1.7 \times 10^{32}$~erg, $M_{1}^{\rm CME} \simeq 9.4 \times 10^{15}$~g) within our M-dwarf simulation. The stellar surface is color coded (purple-green) by the radial magnetic field driving the ambient AWSoM solution. A secondary color scale (magenta-yellow) denotes the density contrast, $n(t)/n^{\rm SS}$, which is used to trace the CME front by the indicated iso-surface. The nominal shock condition, calculated from the Alfv\'enic Mach number of the CME front ($M_{\rm A}^{\rm CME}$,~Eq.~\ref{eq:1}), is encoded simultaneously by the size of the scatter distribution (spheres) and by a tertiary color scale (cyan-red). The field of view is 32\,$R_{\bigstar}$ with a set of selected large-scale magnetic field lines in gray.} 
\label{fig_3}
\end{figure*}

\begin{figure*}[!t]
\centering
\includegraphics[trim=0.0cm 0.1cm 0.1cm 0.0cm, clip=true,width=0.995\textwidth]{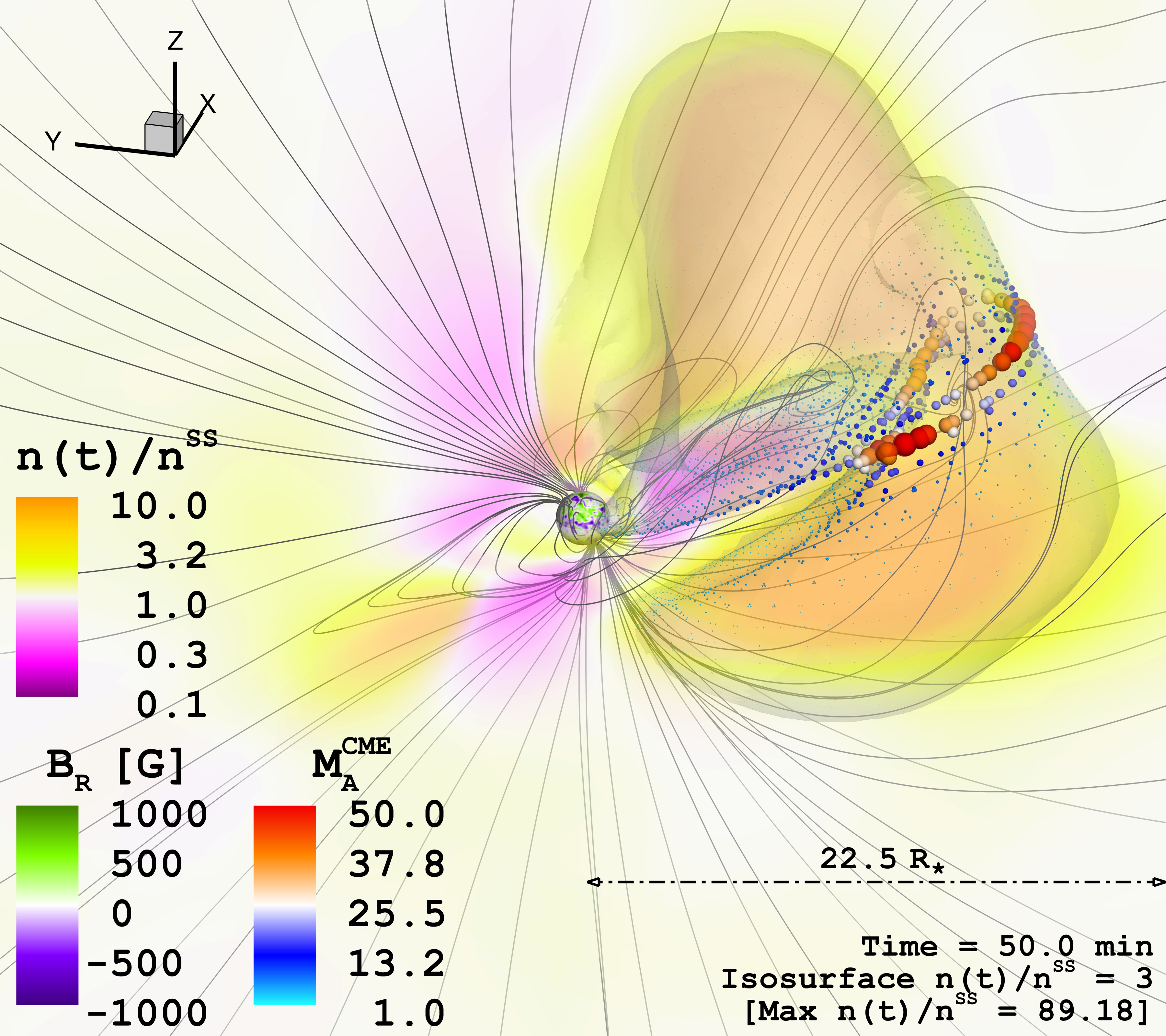}
\caption{Snapshot during the temporal evolution of the partially confined CME ($E_{\rm B,2}^{\rm FR} \simeq 2.0 \times 10^{35}$~erg, $E_{\rm K,2}^{\rm CME} \simeq 3.2 \times 10^{31}$~erg, $M_{2}^{\rm CME} \simeq 4.0 \times 10^{15}$~g) within our M-dwarf simulation. See caption of Fig.~\ref{fig_3}. Note the difference in field of view (45\,$R_{\bigstar}$) and timestamp ($50$~min) in this case.}
\label{fig_4}
\end{figure*}

The influence from a stronger large-scale magnetic field can be clearly seen in the stellar $V_{\rm A}$ distributions (middle and bottom panels of Fig.~\ref{fig_1}). A dipolar geometry is established, with open-field magnetic polar regions showing large $V_{\rm A}$ values which gradually decrease towards the magnetic equator, where the density increases and the field strength decreases (see also Fig.~\ref{fig_2}, middle and bottom panels). Despite the large densities, the strong and ubiquitous small-scale field present in our M-dwarf model increases $V_{\rm A}$ close to the surface. As with the solar case, local minima in $V_{\rm A}$ occur at different positions and heights in the stellar corona. These are nearly absent in our Sun-like simulation, where the large-scale magnetic field dominates the surface distribution. As the large-scale field is weaker in the Sun-like case, $V_{\rm A}$ decays more rapidly with distance compared to the M-dwarf solution. Still, this is only true when evaluated on an equivalent latitude with respect to the large-scale magnetic field dipolar distribution.
  
Finally, it is worth noting that along the current sheet ($\sim0^{\circ}$ in the Sun-like case, $\sim30^{\circ}$ in the M-dwarf model), the Alfv\'en and stellar wind speeds are relatively small (i.e.,~$V_{\rm A}~<~1000$~km~s$^{-1}$, $V^{\rm SW}~\le~100$~km~s$^{-1}$), whereas for other latitudes both quantities increase rapidly (particularly $V_{\rm A}$). As mentioned earlier, this will have important consequences on where in the stellar corona the conditions are more favorable for the escaping CMEs\,---or only certain regions of the expanding structure---\,to become super Alfv\'enic. This also clearly shows the importance of the geometry of the large-scale magnetic field and the need of 3D stellar wind/corona descriptions, neither of which are properly captured in simpler 1D, or even 2D rotationally symmetric, models. The importance the bimodal solar wind to the promotion of CME-driven shock formation has been shown in \citetads{2005ApJ...622.1225M}.

\vspace{1cm}
\setcounter{page}{6}
\subsection{CME Evolution: Magnetic Suppression and Alfv\'enic Regimes}\label{sec:td}

\noindent We now consider the results from the time-dependent CME simulations taking place in our Proxima-like corona and stellar wind environment. As mentioned earlier, two similar TD flux-rope eruptions, only differing by nearly a factor of $2$ in magnetic energy, serve to generate CME events in the regimes of weak (Fig.~\ref{fig_3}) and partial (Fig.~\ref{fig_4}) large-scale magnetic field confinement. 

\begin{figure*}[!t]
\centering
\includegraphics[trim=0.25cm 9.5cm 0.5cm 3.5cm, clip=true,scale=0.3115]{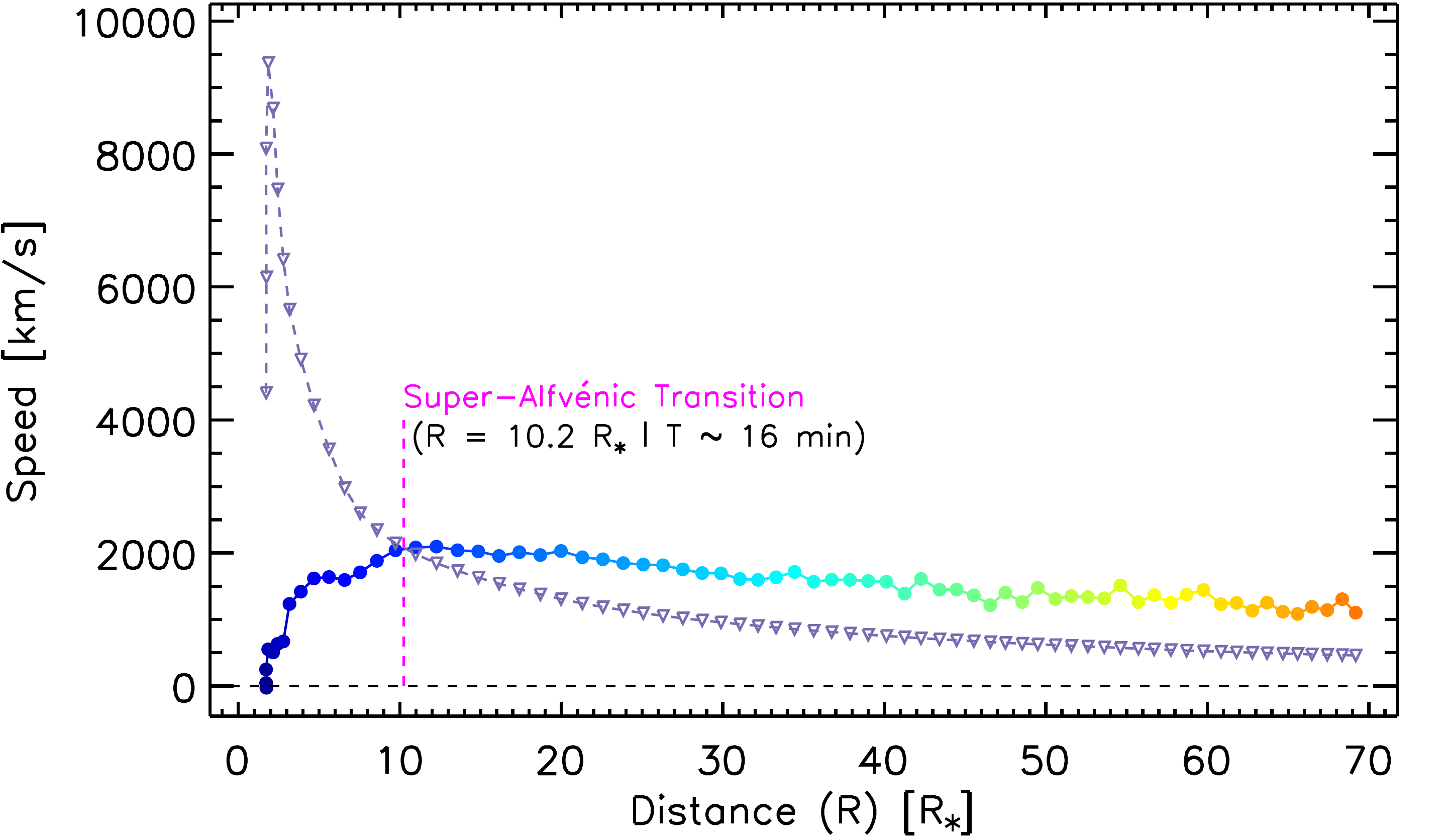}
\includegraphics[trim=1.4cm 9.5cm 0.25cm 3.5cm, clip=true,scale=0.3115]{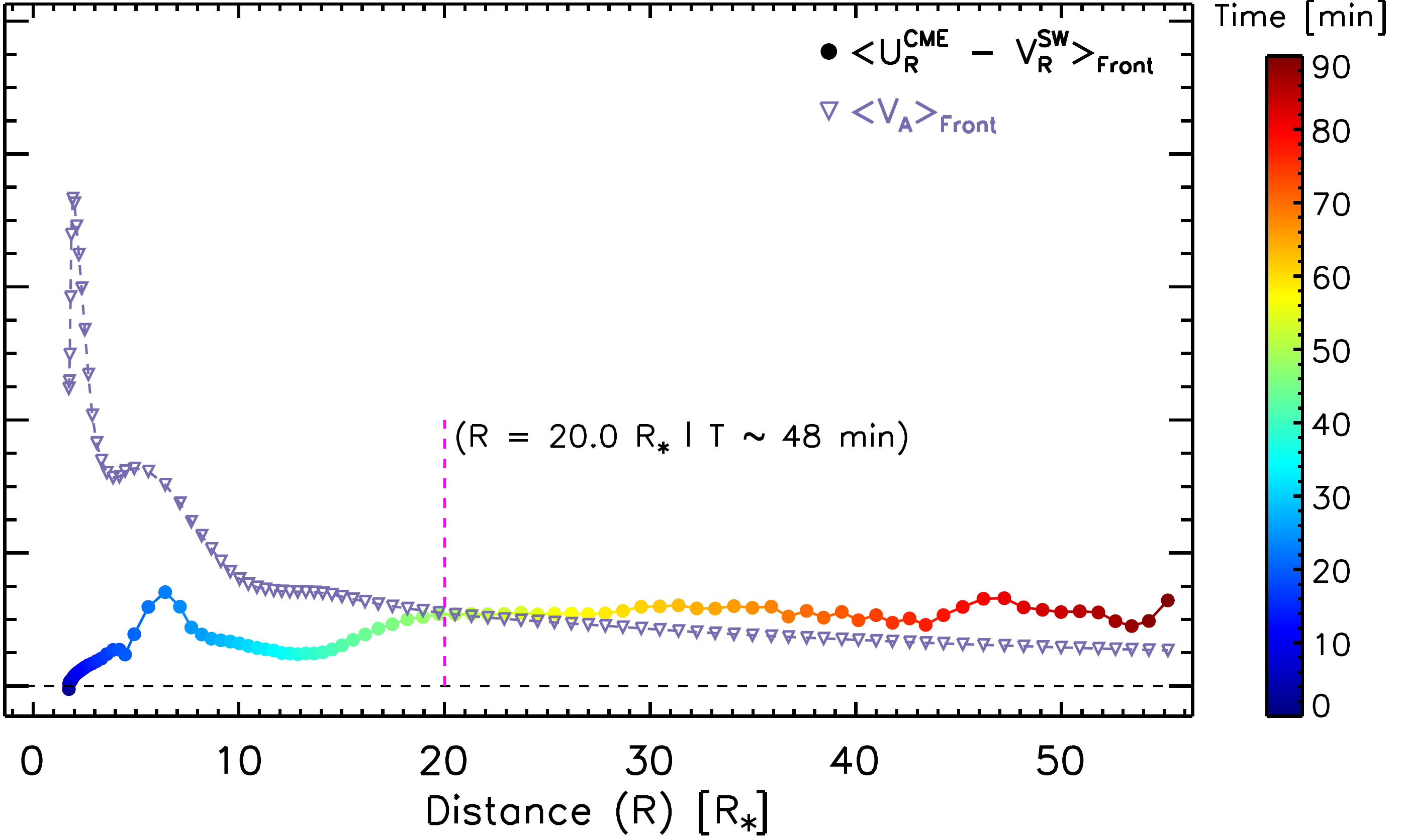}
\caption{Global Alfv\'enic regimes as a function of distance in our simulated M-dwarf CME events (\textit{Left}: weakly suppressed, Fig.~\ref{fig_3}; \textit{Right}: partially suppressed, Fig.~\ref{fig_4}). In circles, and color-coded by simulation time, is the radial CME speed in the stellar wind frame $\left<U_{\rm R}^{\rm CME} - V_{\rm R}^{\rm SW}\right>$, spatially averaged over the expanding front. The mean Alfv\'en speed values $\left<V_{\rm A}\right>$, computed over the same spatial locations, are indicated by downward triangles. The nominal transition from sub-Alfv\'enic, $\left<M_{\rm A}^{\rm CME}\right> < 1$, to super-Alfv\'enic, $\left<M_{\rm A}^{\rm CME}\right> > 1$, is indicated. The spike in speed at a distance of $\sim$$6.5~R_{\bigstar}$ in the right panel is due to the fragmentation of the CME, where the average speed is dominated by the outermost small fragments escaping the large-scale confinement (see Alvarado-G\'omez et al. \citeyearads{2019arXiv191212314A}, \citeyearads{2019ApJ...884L..13A}).}
\label{fig_5}
\end{figure*}

Following \citetads{2019ApJ...884L..13A}, we employ a density contrast $n(t)/n^{\rm SS} = 3.0$ (with $n^{\rm SS}$ as the pre-eruption local density value) to identify and trace the CME front\footnote[11]{This threshold is usually not met during the first $1-2$ minutes of evolution. In those cases, we consider instead $60$\% of the maximum value achieved in $n(t)/n^{\rm SS}$.}. We use the positions of each point on this time-evolving iso-surface to calculate the maximum radial velocity of the CME ($U_{\rm R}^{\rm CME}$), and to extract the steady-state pre-CME stellar wind ($V_{\rm R}^{\rm SW}$) and Alfv\'en ($V_{\rm A}$) speeds on the same locations in each time-step. As described before, this is necessary as the CME motion needs to be transformed to the stellar wind frame in order check the nominal Alfv\'en shock condition:

\begin{equation}\label{eq:1}
\dfrac{(U_{\rm R}^{\rm CME} - V_{\rm R}^{\rm SW})}{V_{\rm A}} \equiv M_{\rm A}^{\rm CME} > 1\mbox{\,.} 
\end{equation}

\noindent Additionally, by integrating over the volume enclosed by the expanding front, taking into account the local escape velocity\footnote[12]{Calculated as $v_{\rm esc} = \sqrt{2GM_{*}/H}$, where $G$ is the gravitational constant and $H$ the front height from the stellar surface.}, we compute the mass ($M^{\rm CME}$) and kinetic energy ($E_{\rm K}^{\rm CME}$) of each CME event. This procedure yields $M_{1}^{\rm CME} \simeq 9.4 \times 10^{15}$~g, $E_{\rm K,1}^{\rm CME}~\simeq~1.7 \times 10^{32}$~erg for the weakly suppressed CME (Fig.~\ref{fig_3}), and a partially suppressed eruption with $M_{2}^{\rm CME} \simeq 4.0 \times 10^{15}$~g, $E_{\rm K,2}^{\rm CME} \simeq 3.2 \times 10^{31}$~erg (Fig.~\ref{fig_4}). 

Figures~\ref{fig_3} and \ref{fig_4} also include a visualization of the emerging shock regions through a distribution of spheres with size and color normalized by $M_{\rm A}^{\rm CME}$. Despite the difference in magnetic energy\,---and therefore in the confinement imposed by the large-scale field---\, both events are able to generate shocks in the corona. As expected from the $V_{\rm A}$ distribution (Sect.~\ref{sec:ss}), in both cases the super-Alfv\'enic region of the CME appears close to the current sheet. Relatively high $M_{\rm A}^{\rm CME}$ values appear locally (a few tens), much larger than in solar observations (see e.g.~\citeads{2020A&A...633A..56M}). Still, most of the perturbation front remains sub-Alfv\'enic as the eruption expands. These result uniquely depends on the 3D setup of the simulations and would not be found in a 1D case.     

One significant difference between the simulated CME events appears in the height with respect to the stellar surface of the shock formation region (note that the field of view and timestamp in Figs~\ref{fig_3} and \ref{fig_4} are different). All the other parameters of the TD flux-ropes being equal, this responds to the available magnetic energy to power the eruption, which in turn determines the relative importance of the magnetic suppression on the emerging eruption properties such as the CME speed (\citeads{2018ApJ...862...93A}).

To better illustrate this, Fig.~\ref{fig_5} shows averages over the CME front of the  radial speed in the stellar wind frame  $\left<U_{\rm R}^{\rm CME} - V_{\rm R}^{\rm SW}\right>$, and the local Alfv\'en speed $\left<V_{\rm A}\right>$, as a function of distance in both events. The resulting global behavior shows that the sub- to super-Alfv\'enic transition, indicative of the shock formation region, occurs at several stellar radii of height for both events (roughly at $10$~$R_{\bigstar}$ and $20$ $R_{\bigstar}$ for the events in Fig.~\ref{fig_3} and \ref{fig_4}, respectively). As discussed in the following section, this will have important consequences for any Type~II bursts radio signatures induced by these shocks, and their detectability in the stellar regime with current instrumentation.

Finally, given the very large magnetic fields in the upper corona, temperature effects due to heating at radius $< 10$~$R_{\bigstar}$ can be neglected. We have verified that along the current sheet in the M-dwarf simulation the wind temperature is $T < 3$~MK. The resulting sound speed, for an hydrogen ideal gas, is $\sim 200$~km~s$^{-1} \ll V_{\rm A}$, so that the fast magnetosonic Mach number can be approximated with $M_{\rm A}^{\rm CME}$ within $< 1 \%$. Local larger temperature would push further out the distance of transition to super-Alfv\'enic CME speed by a modest amount.  

\section{Discussion}\label{sec:discussion}

\noindent Solar type II radio bursts are typically divided by their associated wavelength or starting frequency (see \citeads{2017Ap.....60..213S} and references therein). Coronal type II radio bursts manifest at decimeter to metric wavelengths (MHz range), and interplanetary (IP) type II radio bursts appear at decametric to kilometric wavelengths (kHz range). One fundamental aspect related to their detection is the fact that the ionosphere impedes the transmission of radio waves with frequencies below $\sim$$10$~MHz (cutoff frequency\footnote[13]{This is a nominal average value which, among other factors, has  diurnal, seasonal, and solar activity-related variations (see~\citeads{2018assi.book.....Y}).}). Therefore, only coronal type~II bursts are accessible from ground-based instrumentation, which is also the sole possibility for their search in the context of stellar CMEs (see \citeads{2017PhDT.........8V} and references therein). 

The type II radio burst division is clearly motivated by the expected shock formation region. Still, several solar events display emission in the entire radio domain (m-to-km type II bursts). \citetads{2005JGRA..11012S07G} studied the properties of m-to-km type II radio bursts and their driving CMEs. This statistical analysis revealed that the majority of such radio events form close to the solar surface (i.e., below $3~R_{\odot}$ of height), and that the kinetic energy of the CME controls the life time of the radio emission (i.e., the range of frequencies covered by a given event). When the sample is restricted to coronal type II radio bursts alone, \citetads{2005JGRA..11012S07G} reports that the average shock formation region is even lower in height ($< 2~R_{\odot}$; see also \citeads{2012ApJ...752..107R}, \citeads{2013AdSpR..51.1981G}). 

\begin{figure}[!t]
\centering
\includegraphics[trim=1.3cm 0.0cm 0.98cm 0.0cm, clip=true,width=0.48\textwidth]{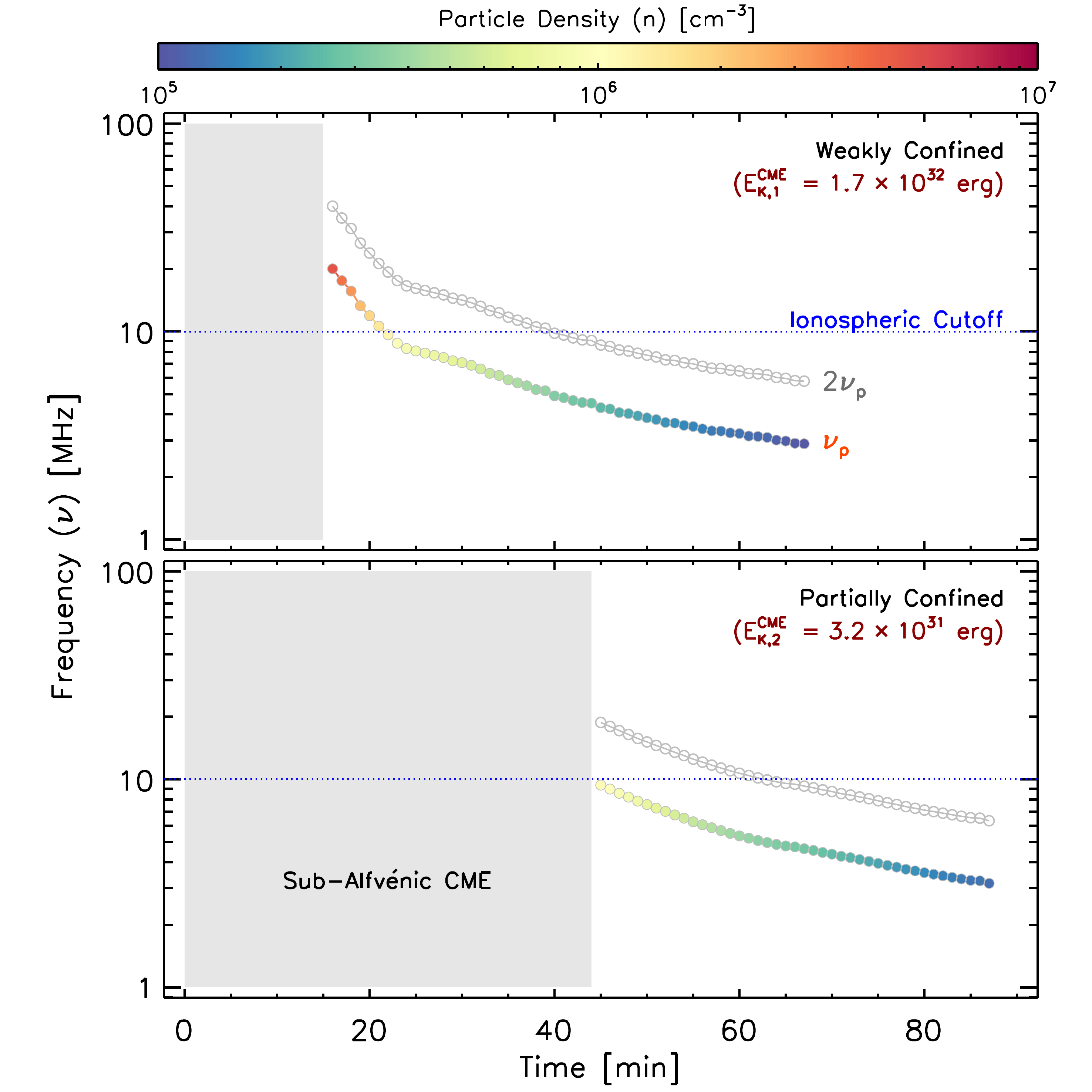}
\caption{Temporal evolution of the expected type II radio burst frequencies associated with the shock regions in our simulated M-dwarf CMEs (\textit{Top}: Weakly confined |Fig.~\ref{fig_3}; \textit{Bottom}: Partially confined |Fig.~\ref{fig_4}). The fundamental ($\nu_{\rm p} \simeq 8980\sqrt{n}$~[Hz]) and first harmonic ($2\nu_{\rm p}$) of the plasma frequency are included. The associated $M_{\rm A}^{\rm CME}-$weighted mean plasma density is indicated by the color scale. Gray regions denote the intervals in which each CME is within the sub-Alfv\'enic regime (see Fig.~\ref{fig_5}).}
\label{fig_6}
\end{figure}
  
Our analysis indicates that the shocks generated by the simulated M-dwarf CME events are pushed much farther out (see Fig.~\ref{fig_5}). At such distances, the coronal densities have decreased substantially compared to the standard formation region of solar type II bursts, shifting their frequencies close to, or below, the ionospheric cutoff. This is presented in Fig.~\ref{fig_6}, where the expected fundamental, $\nu_{\rm p} \simeq 8980\sqrt{n}$ [Hz], and first harmonic, $2\nu_{\rm p}$, of the plasma frequency are shown as a function of time. As regions with stronger shocks are expected to contribute more to the global type II emission, $M_{\rm A}^{\rm CME}-$weighted average densities are considered for this calculation.

Nevertheless, Fig.~\ref{fig_6} shows that our simulated M-dwarf CMEs are not entirely radio-quiet. The expected fundamental and harmonic type II burst frequency drifts, generated by the strongest eruption considered here (Fig.~\ref{fig_6}, top panel), remain above the ionospheric cutoff for $\sim$\,10 min and $\sim$\,30 min, respectively. With approximately $\sim$\,80\% less kinetic energy, only a $\sim$\,15 min harmonic lane clears this threshold in the weaker CME event (Fig.~\ref{fig_6}, bottom panel).

We now estimate whether such stellar type II radio bursts could be detected with ground-based instruments like the LOw Frequency ARray (LOFAR, \citeads{2013A&A...556A...2V}). The strongest solar type II radio bursts reach spectral fluxes up to $10^8~\rm{Jy}$ (\citeads{2016GeoRL..43...50S}). If we assume that this is also representative for M-dwarfs, that corresponds to $1.4~\rm{mJy}$ from Proxima Centauri's distance of $1.3~\rm{pc}$, although this specific object in the Southern sky is not visible to LOFAR with a core location at a geographic latitude of 53 degrees North. LOFAR provides online tools for sensitivity estimates\footnote[14]{\url{https://support.astron.nl/ImageNoiseCalculator/sens.php}}. For an observing frequency around $30~\rm{MHz}$, a maximum burst duration of $30~\rm{min}$, and a typical instantaneous burst bandwidth of $20~\rm{MHz}$ (\citeads{2019NatAs...3..452M}), this leads to a sensitivity of just $5~\rm{mJy}$. And this number has to be treated with caution, since this estimate does not consider effects like calibration errors, ionospheric conditions, elevation of the source, and errors in beam models. The application of a factor of 5 is advised. 

So it has to be concluded that even in the best case, i.e.~harmonic emission from the weakly confined case in the upper panel of Fig.~\ref{fig_6}, with a total flux equal to the maximum solar value, CME-related type II radio burst emission cannot be observed by LOFAR even for the nearest M-dwarfs. The upcoming Square Kilometre Array (SKA), with its eponymous collecting area, could provide the necessary sensitivity and also geographic location for observations of Proxima Centauri. However, the lowest frequency band of SKA just starts at $50~\rm{MHz}$ (\citeads{2019AdSpR..63.1404N}). From Fig.~\ref{fig_6} it follows that there is no radio emission from frequencies above $50~\rm{MHz}$, as the sources of both fundamental and harmonic emission would be located in the region where the CME is still sub-Alfv\'enic.

In agreement with the results of \citetads{2019ApJ...873....1M}, we find then that the ground-based detection of stellar CMEs via type II burst emission would be greatly hampered by the combined effect from magnetic suppression and large $V_{\rm A}$ values in the corona. It is worth noting that our Proxima Centauri background steady-state model already provides a best-case scenario: a lower bound on the mean surface field strength ($\sim\,$450~G; \citeads{2008A&A...489L..45R}), highest stellar wind density allowed by observations ($\dot{\rm M} \simeq 0.3$~$\dot{\rm M}_{\odot}$; \citeads{2001ApJ...547L..49W}), and a CME shock trajectory following the current sheet (i.e.,~global minimum of $V_{\rm A}$ and low $V^{\rm SW}$; see bottom panels of Figs.~\ref{fig_1}~and~\ref{fig_2}). Still, our analysis shows that the required CME speed of $10$\% the speed of light, suggested by \citetads{2019ApJ...873....1M} for Proxima-like surface magnetic fields, is overestimated. This most likely reflects the lower dimensionality and much simpler coronal model assumed in their study. 

In terms of our simulated eruptions, the kinetic energies appear conservative with respect to the range determined for the best CME candidate observed in Proxima Centauri so far ($1 \times 10^{29}$~erg $< E_{\rm K}^{\rm CME} < 4 \times 10^{34}$~erg; \citeads{2019ApJ...877..105M}). Leaving aside considerations on occurrence rate, more energetic CMEs might be able to shock higher density regions closer to the stellar surface, increasing the radio frequency of the associated type II bursts. However, kinetic energies of CMEs appear correlated with small-scale surface magnetic flux in solar observations (e.g.,~\citeads{2017ApJ...834...56T}, \citeads{2020ApJ...889..104S}). If such a  relation holds for stellar CMEs, it implies that $V_{\rm A}$ could also increase locally for stronger eruptions, creating again unfavorable conditions for the generation of shocks in the low corona. Still, this effect might be secondary for certain large-scale magnetic field strengths (i.e., reduced CME suppression). Future investigation will be pursued in this direction, expanding the parameter space to additional spectral types, surface magnetic field configurations, stellar wind properties, and CME eruption models.  

\section{Summary}\label{sec:summary}

\noindent Continuing our numerical investigation on stellar CMEs, we have considered here the expected connection between these eruptive phenomena and the generation of type II radio bursts. 

Using physics-based 3D MHD corona and stellar wind models, we compared the Alfv\'en speed distribution for the Sun, a young Sun-like star, and a moderately active M-dwarf. We examined the regions where the Alfv\'en and stellar wind speeds provide the most favorable conditions for the generation of shocks in the corona. Furthermore, employing a state-of-the-art flux-rope eruption model, we simulated admissible CME events occurring in the archetypical star Proxima Centauri. We considered two eruptions representative of the regimes over which the ejected plasma is able to escape the large-scale magnetic field (weak and partial confinement). We showed that these eruptions are able to generate local strong shocks (i.e., high Alfv\'enic Mach number) in the vicinity of the astrospheric current sheet, which may lead to efficient acceleration of charged particles not due to magnetic reconnection. 

The analysis of the global behavior in each CME event revealed that the shock formation region is pushed outwards compared to the average location observed in the Sun. From this, it follows that the associated type II radio burst frequencies would be shifted to lower values as the kinetic energy of the CME decreases. This poses a challenge for their detection from the ground, as in some cases their radio emission would lie very close to, or below, the ionospheric frequency cutoff. Nevertheless, extreme events might be able to more rapidly overcome the large-scale magnetic field suppression, decreasing the shock formation height and yielding amenable type II burst frequencies for current and future ground-based facilities. 

\acknowledgments
\noindent We would like to thank the referee for constructive feedback. J.D.A.G. was supported by Chandra GO5-16021X and HST GO-15326 grants.  S.P.M. and O.C. were supported by NASA Living with a Star grant number NNX16AC11G. J.J.D. was funded by NASA contract NAS8-03060 to the Chandra X-ray Center and thanks the director, Belinda Wilkes, for continuing advice and support. This work was carried out using the SWMF/BATSRUS tools developed at the University of Michigan Center for Space Environment Modeling (CSEM) and made available through the NASA Community Coordinated Modeling Center (CCMC). Resources supporting this work were provided by the NASA High-End Computing (HEC) Program through the NASA Advanced Supercomputing (NAS) Division at Ames Research Center. Simulations were performed on NASA's Pleiades cluster under award SMD-17-1330.

%

\facilities{NASA Pleiades Supercomputer}


\software{SWMF \citepads{2018LRSP...15....4G}}

\bibliographystyle{aasjournal}
\bibliography{Biblio}

\begin{thebibliography}{}
\expandafter\ifx\csname natexlab\endcsname\relax\def\natexlab#1{#1}\fi
\providecommand{\url}[1]{\href{#1}{#1}}

\bibitem[{{Alvarado-G{\'o}mez} {et~al.}(2018){Alvarado-G{\'o}mez}, {Drake},
  {Cohen}, {Moschou}, \& {Garraffo}}]{2018ApJ...862...93A}
{Alvarado-G{\'o}mez}, J.~D., {Drake}, J.~J., {Cohen}, O., {Moschou}, S.~P., \&
  {Garraffo}, C. 2018, \apj, 862, 93

\bibitem[{{Alvarado-G{\'o}mez}
  {et~al.}(2019{\natexlab{a}}){Alvarado-G{\'o}mez}, {Drake}, {Garraffo},
  {Moschou}, {Cohen}, {Yadav}, \& {Fraschetti}}]{2019arXiv191212314A}
{Alvarado-G{\'o}mez}, J.~D., {Drake}, J.~J., {Garraffo}, C., {et~al.}
  2019{\natexlab{a}}, arXiv e-prints, arXiv:1912.12314

\bibitem[{{Alvarado-G{\'o}mez}
  {et~al.}(2019{\natexlab{b}}){Alvarado-G{\'o}mez}, {Drake}, {Moschou},
  {Garraffo}, {Cohen}, {NASA LWS Focus Science Team: The Solar-Stellar
  Connection}, {Yadav}, \& {Fraschetti}}]{2019ApJ...884L..13A}
{Alvarado-G{\'o}mez}, J.~D., {Drake}, J.~J., {Moschou}, S.~P., {et~al.}
  2019{\natexlab{b}}, \apjl, 884, L13

\bibitem[{{Alvarado-G{\'o}mez}
  {et~al.}(2019{\natexlab{c}}){Alvarado-G{\'o}mez}, {Garraffo}, {Drake},
  {Brown}, {Oishi}, {Moschou}, \& {Cohen}}]{2019ApJ...875L..12A}
{Alvarado-G{\'o}mez}, J.~D., {Garraffo}, C., {Drake}, J.~J., {et~al.}
  2019{\natexlab{c}}, \apj, 875, L12

\bibitem[{{Argiroffi} {et~al.}(2019){Argiroffi}, {Reale}, {Drake},
  {Ciaravella}, {Testa}, {Bonito}, {Miceli}, {Orlando}, \&
  {Peres}}]{2019NatAs.tmp..328A}
{Argiroffi}, C., {Reale}, F., {Drake}, J.~J., {et~al.} 2019, Nature Astronomy,
  328

\bibitem[{{Benz}(2017)}]{2017LRSP...14....2B}
{Benz}, A.~O. 2017, Living Reviews in Solar Physics, 14, 2

\bibitem[{{Bilenko}(2018)}]{2018Ge&Ae..58..989B}
{Bilenko}, I.~A. 2018, Geomagnetism and Aeronomy, 58, 989

\bibitem[{{Cairns} {et~al.}(2003){Cairns}, {Knock}, {Robinson}, \&
  {Kuncic}}]{2003SSRv..107...27C}
{Cairns}, I.~H., {Knock}, S.~A., {Robinson}, P.~A., \& {Kuncic}, Z. 2003, \ssr,
  107, 27

\bibitem[{{Caramazza} {et~al.}(2007){Caramazza}, {Flaccomio}, {Micela},
  {Reale}, {Wolk}, \& {Feigelson}}]{2007A&A...471..645C}
{Caramazza}, M., {Flaccomio}, E., {Micela}, G., {et~al.} 2007, \aap, 471, 645

\bibitem[{{Cherenkov} {et~al.}(2017){Cherenkov}, {Bisikalo}, {Fossati}, \&
  {M{\"o}stl}}]{2017ApJ...846...31C}
{Cherenkov}, A., {Bisikalo}, D., {Fossati}, L., \& {M{\"o}stl}, C. 2017, \apj,
  846, doi:10.3847/1538-4357/aa82b2

\bibitem[{{Cohen} {et~al.}(2017){Cohen}, {Yadav}, {Garraffo}, {Saar}, {Wolk},
  {Kashyap}, {Drake}, \& {Pillitteri}}]{2017ApJ...834...14C}
{Cohen}, O., {Yadav}, R., {Garraffo}, C., {et~al.} 2017, \apj, 834, 14

\bibitem[{{Collins} {et~al.}(2017){Collins}, {Jones}, \&
  {Barnes}}]{2017A&A...602A..48C}
{Collins}, J.~M., {Jones}, H. R.~A., \& {Barnes}, J.~R. 2017, \aap, 602, A48

\bibitem[{{Compagnino} {et~al.}(2017){Compagnino}, {Romano}, \&
  {Zuccarello}}]{2017SoPh..292....5C}
{Compagnino}, A., {Romano}, P., \& {Zuccarello}, F. 2017, \solphys, 292, 5

\bibitem[{{Cranmer}(2017)}]{2017ApJ...840..114C}
{Cranmer}, S.~R. 2017, \apj, 840, 114

\bibitem[{{Crosley} \& {Osten}(2018{\natexlab{a}})}]{2018ApJ...856...39C}
{Crosley}, M.~K., \& {Osten}, R.~A. 2018{\natexlab{a}}, \apj, 856,
  doi:10.3847/1538-4357/aaaec2

\bibitem[{{Crosley} \& {Osten}(2018{\natexlab{b}})}]{2018ApJ...862..113C}
---. 2018{\natexlab{b}}, \apj, 862, 113

\bibitem[{{Crosley} {et~al.}(2016){Crosley}, {Osten}, {Broderick}, {Corbel},
  {Eisl{\"o}ffel}, {Grie{\ss}meier}, {van Leeuwen}, {Rowlinson}, {Zarka}, \&
  {Norman}}]{2016ApJ...830...24C}
{Crosley}, M.~K., {Osten}, R.~A., {Broderick}, J.~W., {et~al.} 2016, \apj, 830,
  24

\bibitem[{{Davenport}(2016)}]{2016ApJ...829...23D}
{Davenport}, J.~R.~A. 2016, \apj, 829, 23

\bibitem[{{Davenport} {et~al.}(2019){Davenport}, {Covey}, {Clarke}, {Boeck},
  {Cornet}, \& {Hawley}}]{2019ApJ...871..241D}
{Davenport}, J. R.~A., {Covey}, K.~R., {Clarke}, R.~W., {et~al.} 2019, \apj,
  871, 241

\bibitem[{{DeRosa} \& {Barnes}(2018)}]{2018ApJ...861..131D}
{DeRosa}, M.~L., \& {Barnes}, G. 2018, \apj, 861, 131

\bibitem[{{Donati}(2011)}]{2011IAUS..271...23D}
{Donati}, J.-F. 2011, in IAU Symposium, Vol. 271, Astrophysical Dynamics: From
  Stars to Galaxies, ed. N.~H. {Brummell}, A.~S. {Brun}, M.~S. {Miesch}, \&
  Y.~{Ponty}, 23--31

\bibitem[{{Drake} {et~al.}(2016){Drake}, {Cohen}, {Garraffo}, \&
  {Kashyap}}]{2016IAUS..320..196D}
{Drake}, J.~J., {Cohen}, O., {Garraffo}, C., \& {Kashyap}, V. 2016, in IAU
  Symposium, Vol. 320, Solar and Stellar Flares and their Effects on Planets,
  ed. A.~G. {Kosovichev}, S.~L. {Hawley}, \& P.~{Heinzel}, 196--201

\bibitem[{{Drake} {et~al.}(2013){Drake}, {Cohen}, {Yashiro}, \&
  {Gopalswamy}}]{2013ApJ...764..170D}
{Drake}, J.~J., {Cohen}, O., {Yashiro}, S., \& {Gopalswamy}, N. 2013, \apj,
  764, 170

\bibitem[{{Emslie} {et~al.}(2012){Emslie}, {Dennis}, {Shih}, {Chamberlin},
  {Mewaldt}, {Moore}, {Share}, {Vourlidas}, \& {Welsch}}]{2012ApJ...759...71E}
{Emslie}, A.~G., {Dennis}, B.~R., {Shih}, A.~Y., {et~al.} 2012, \apj, 759, 71

\bibitem[{{Fraschetti} {et~al.}(2018){Fraschetti}, {Drake}, {Cohen}, \&
  {Garraffo}}]{2018ApJ...853..112F}
{Fraschetti}, F., {Drake}, J.~J., {Cohen}, O., \& {Garraffo}, C. 2018, \apj,
  853, 112

\bibitem[{{Garraffo} {et~al.}(2016{\natexlab{a}}){Garraffo}, {Drake}, \&
  {Cohen}}]{2016A&A...595A.110G}
{Garraffo}, C., {Drake}, J.~J., \& {Cohen}, O. 2016{\natexlab{a}}, \aap, 595,
  A110

\bibitem[{{Garraffo} {et~al.}(2016{\natexlab{b}}){Garraffo}, {Drake}, \&
  {Cohen}}]{2016ApJ...833L...4G}
---. 2016{\natexlab{b}}, \apjl, 833, L4

\bibitem[{{Glassgold} {et~al.}(1997){Glassgold}, {Najita}, \&
  {Igea}}]{1997ApJ...480..344G}
{Glassgold}, A.~E., {Najita}, J., \& {Igea}, J. 1997, \apj, 480, 344

\bibitem[{{Gombosi} {et~al.}(2018){Gombosi}, {van der Holst}, {Manchester}, \&
  {Sokolov}}]{2018LRSP...15....4G}
{Gombosi}, T.~I., {van der Holst}, B., {Manchester}, W.~B., \& {Sokolov}, I.~V.
  2018, Living Reviews in Solar Physics, 15, 4

\bibitem[{{Gopalswamy} {et~al.}(2005){Gopalswamy}, {Aguilar-Rodriguez},
  {Yashiro}, {Nunes}, {Kaiser}, \& {Howard}}]{2005JGRA..11012S07G}
{Gopalswamy}, N., {Aguilar-Rodriguez}, E., {Yashiro}, S., {et~al.} 2005,
  Journal of Geophysical Research (Space Physics), 110, A12S07

\bibitem[{{Gopalswamy} {et~al.}(2019){Gopalswamy}, {M{\"a}kel{\"a}}, \&
  {Yashiro}}]{2019arXiv191207370G}
{Gopalswamy}, N., {M{\"a}kel{\"a}}, P., \& {Yashiro}, S. 2019, arXiv e-prints,
  arXiv:1912.07370

\bibitem[{{Gopalswamy} {et~al.}(2009){Gopalswamy}, {Thompson}, {Davila},
  {Kaiser}, {Yashiro}, {M{\"a}kel{\"a}}, {Michalek}, {Bougeret}, \&
  {Howard}}]{2009SoPh..259..227G}
{Gopalswamy}, N., {Thompson}, W.~T., {Davila}, J.~M., {et~al.} 2009, \solphys,
  259, 227

\bibitem[{{Gopalswamy} {et~al.}(2013){Gopalswamy}, {Xie}, {M{\"a}kel{\"a}},
  {Yashiro}, {Akiyama}, {Uddin}, {Srivastava}, {Joshi}, {Chandra}, {Manoharan},
  {Mahalakshmi}, {Dwivedi}, {Jain}, {Awasthi}, {Nitta}, {Aschwand en}, \&
  {Choudhary}}]{2013AdSpR..51.1981G}
{Gopalswamy}, N., {Xie}, H., {M{\"a}kel{\"a}}, P., {et~al.} 2013, Advances in
  Space Research, 51, 1981

\bibitem[{{Guarcello} {et~al.}(2019){Guarcello}, {Micela}, {Sciortino},
  {L{\'o}pez-Santiago}, {Argiroffi}, {Reale}, {Flaccomio},
  {Alvarado-G{\'o}mez}, {Antoniou}, {Drake}, {Pillitteri}, {Rebull}, \&
  {Stauffer}}]{2019A&A...622A.210G}
{Guarcello}, M.~G., {Micela}, G., {Sciortino}, S., {et~al.} 2019, \aap, 622,
  A210

\bibitem[{{Ilin} {et~al.}(2019){Ilin}, {Schmidt}, {Davenport}, \&
  {Strassmeier}}]{2019A&A...622A.133I}
{Ilin}, E., {Schmidt}, S.~J., {Davenport}, J. R.~A., \& {Strassmeier}, K.~G.
  2019, \aap, 622, A133

\bibitem[{{Jeffers} {et~al.}(2017){Jeffers}, {Boro Saikia}, {Barnes}, {Petit},
  {Marsden}, {Jardine}, {Vidotto}, \& {BCool
  Collaboration}}]{2017MNRAS.471L..96J}
{Jeffers}, S.~V., {Boro Saikia}, S., {Barnes}, J.~R., {et~al.} 2017, \mnras,
  471, L96

\bibitem[{{Jeffers} {et~al.}(2014){Jeffers}, {Petit}, {Marsden}, {Morin},
  {Donati}, \& {Folsom}}]{2014A&A...569A..79J}
{Jeffers}, S.~V., {Petit}, P., {Marsden}, S.~C., {et~al.} 2014, \aap, 569, A79

\bibitem[{{Jin} {et~al.}(2017{\natexlab{a}}){Jin}, {Manchester}, {van der
  Holst}, {Sokolov}, {T{\'o}th}, {Vourlidas}, {de Koning}, \&
  {Gombosi}}]{2017ApJ...834..172J}
{Jin}, M., {Manchester}, W.~B., {van der Holst}, B., {et~al.}
  2017{\natexlab{a}}, \apj, 834, 172

\bibitem[{{Jin} {et~al.}(2013){Jin}, {Manchester}, {van der Holst}, {Oran},
  {Sokolov}, {Toth}, {Liu}, {Sun}, \& {Gombosi}}]{2013ApJ...773...50J}
---. 2013, \apj, 773, 50

\bibitem[{{Jin} {et~al.}(2017{\natexlab{b}}){Jin}, {Manchester}, {van der
  Holst}, {Sokolov}, {T{\'o}th}, {Mullinix}, {Taktakishvili}, {Chulaki}, \&
  {Gombosi}}]{2017ApJ...834..173J}
---. 2017{\natexlab{b}}, \apj, 834, 173

\bibitem[{{Kashyap} {et~al.}(2002){Kashyap}, {Drake}, {G{\"u}del}, \&
  {Audard}}]{2002ApJ...580.1118K}
{Kashyap}, V.~L., {Drake}, J.~J., {G{\"u}del}, M., \& {Audard}, M. 2002, \apj,
  580, 1118

\bibitem[{{Kervella} {et~al.}(2017){Kervella}, {Th{\'e}venin}, \&
  {Lovis}}]{2017A&A...598L...7K}
{Kervella}, P., {Th{\'e}venin}, F., \& {Lovis}, C. 2017, \aap, 598, L7

\bibitem[{{Kiraga} \& {Stepien}(2007)}]{2007AcA....57..149K}
{Kiraga}, M., \& {Stepien}, K. 2007, \actaa, 57, 149

\bibitem[{{Kundu}(1965)}]{1965sra..book.....K}
{Kundu}, M.~R. 1965, {Solar radio astronomy}

\bibitem[{{Lammer}(2013)}]{2013oepa.book.....L}
{Lammer}, H. 2013, {Origin and Evolution of Planetary Atmospheres} (Springer
  Berlin Heidelberg), doi:10.1007/978-3-642-32087-3

\bibitem[{{Linsky} \& {Wood}(2014)}]{2014ASTRP...1...43L}
{Linsky}, J.~L., \& {Wood}, B.~E. 2014, ASTRA Proceedings, 1, 43

\bibitem[{{Liu} {et~al.}(2016){Liu}, {Wang}, {Wang}, {Shen}, {Ye}, {Liu},
  {Chen}, {Zhang}, \& {Wang}}]{2016ApJ...826..119L}
{Liu}, L., {Wang}, Y., {Wang}, J., {et~al.} 2016, \apj, 826, 119

\bibitem[{{Loyd} {et~al.}(2018){Loyd}, {France}, {Youngblood}, {Schneider},
  {Brown}, {Hu}, {Segura}, {Linsky}, {Redfield}, {Tian}, {Rugheimer}, {Miguel},
  \& {Froning}}]{2018ApJ...867...71L}
{Loyd}, R.~O.~P., {France}, K., {Youngblood}, A., {et~al.} 2018, \apj, 867, 71

\bibitem[{{Maguire} {et~al.}(2020){Maguire}, {Carley}, {McCauley}, \&
  {Gallagher}}]{2020A&A...633A..56M}
{Maguire}, C.~A., {Carley}, E.~P., {McCauley}, J., \& {Gallagher}, P.~T. 2020,
  \aap, 633, A56

\bibitem[{{Manchester} {et~al.}(2005){Manchester}, {Gombosi}, {De Zeeuw},
  {Sokolov}, {Roussev}, {Powell}, {K{\'o}ta}, {T{\'o}th}, \&
  {Zurbuchen}}]{2005ApJ...622.1225M}
{Manchester}, W.~B., I., {Gombosi}, T.~I., {De Zeeuw}, D.~L., {et~al.} 2005,
  \apj, 622, 1225

\bibitem[{{Manchester} {et~al.}(2008){Manchester}, {Vourlidas}, {T{\'o}th},
  {Lugaz}, {Roussev}, {Sokolov}, {Gombosi}, {De Zeeuw}, \&
  {Opher}}]{2008ApJ...684.1448M}
{Manchester}, IV, W.~B., {Vourlidas}, A., {T{\'o}th}, G., {et~al.} 2008, \apj,
  684, 1448

\bibitem[{{Mann} {et~al.}(2003){Mann}, {Klassen}, {Aurass}, \&
  {Classen}}]{2003A&A...400..329M}
{Mann}, G., {Klassen}, A., {Aurass}, H., \& {Classen}, H.~T. 2003, \aap, 400,
  329

\bibitem[{{Micela}(2018)}]{2018haex.bookE..19M}
{Micela}, G. 2018, {Stellar Coronal Activity and Its Impact on Planets}
  (Springer International Publishing AG), 19

\bibitem[{{Morgenthaler} {et~al.}(2012){Morgenthaler}, {Petit}, {Saar},
  {Solanki}, {Morin}, {Marsden}, {Auri{\`e}re}, {Dintrans}, {Fares}, {Gastine},
  {Lanoux}, {Ligni{\`e}res}, {Paletou}, {Ram{\'{\i}}rez V{\'e}lez},
  {Th{\'e}ado}, \& {Van Grootel}}]{2012A&A...540A.138M}
{Morgenthaler}, A., {Petit}, P., {Saar}, S., {et~al.} 2012, \aap, 540, A138

\bibitem[{{Morosan} {et~al.}(2019){Morosan}, {Carley}, {Hayes}, {Murray},
  {Zucca}, {Fallows}, {McCauley}, {Kilpua}, {Mann}, {Vocks}, \&
  {Gallagher}}]{2019NatAs...3..452M}
{Morosan}, D.~E., {Carley}, E.~P., {Hayes}, L.~A., {et~al.} 2019, Nature
  Astronomy, 3, 452

\bibitem[{{Moschou} {et~al.}(2017){Moschou}, {Drake}, {Cohen},
  {Alvarado-Gomez}, \& {Garraffo}}]{2017ApJ...850..191M}
{Moschou}, S.-P., {Drake}, J.~J., {Cohen}, O., {Alvarado-Gomez}, J.~D., \&
  {Garraffo}, C. 2017, \apj, 850, 191

\bibitem[{{Moschou} {et~al.}(2019){Moschou}, {Drake}, {Cohen},
  {Alvarado-G{\'o}mez}, {Garraffo}, \& {Fraschetti}}]{2019ApJ...877..105M}
{Moschou}, S.-P., {Drake}, J.~J., {Cohen}, O., {et~al.} 2019, \apj, 877, 105

\bibitem[{{Mullan} \& {Bais}(2018)}]{2018ApJ...865..101M}
{Mullan}, D.~J., \& {Bais}, H.~P. 2018, \apj, 865, 101

\bibitem[{{Mullan} \& {Paudel}(2019)}]{2019ApJ...873....1M}
{Mullan}, D.~J., \& {Paudel}, R.~R. 2019, \apj, 873, 1

\bibitem[{{Nindos} {et~al.}(2019){Nindos}, {Kontar}, \&
  {Oberoi}}]{2019AdSpR..63.1404N}
{Nindos}, A., {Kontar}, E.~P., \& {Oberoi}, D. 2019, Advances in Space
  Research, 63, 1404

\bibitem[{{Nutzman} \& {Charbonneau}(2008)}]{2008PASP..120..317N}
{Nutzman}, P., \& {Charbonneau}, D. 2008, \pasp, 120, 317

\bibitem[{{Odert} {et~al.}(2017){Odert}, {Leitzinger}, {Hanslmeier}, \&
  {Lammer}}]{2017MNRAS.472..876O}
{Odert}, P., {Leitzinger}, M., {Hanslmeier}, A., \& {Lammer}, H. 2017, \mnras,
  472, 876

\bibitem[{{Oran} {et~al.}(2017){Oran}, {Landi}, {van der Holst}, {Sokolov}, \&
  {Gombosi}}]{2017ApJ...845...98O}
{Oran}, R., {Landi}, E., {van der Holst}, B., {Sokolov}, I.~V., \& {Gombosi},
  T.~I. 2017, \apj, 845, 98

\bibitem[{{Osten}(2016)}]{2016hasa.book...23O}
{Osten}, R. 2016, {Solar explosive activity throughout the evolution of the
  solar system}, 23--55

\bibitem[{{Osten} \& {Wolk}(2017)}]{2017IAUS..328..243O}
{Osten}, R.~A., \& {Wolk}, S.~J. 2017, in IAU Symposium, Vol. 328, Living
  Around Active Stars, ed. D.~{Nandy}, A.~{Valio}, \& P.~{Petit}, 243--251

\bibitem[{{Pevtsov} {et~al.}(2003){Pevtsov}, {Fisher}, {Acton}, {Longcope},
  {Johns-Krull}, {Kankelborg}, \& {Metcalf}}]{2003ApJ...598.1387P}
{Pevtsov}, A.~A., {Fisher}, G.~H., {Acton}, L.~W., {et~al.} 2003, \apj, 598,
  1387

\bibitem[{{Pick} {et~al.}(2006){Pick}, {Forbes}, {Mann}, {Cane}, {Chen},
  {Ciaravella}, {Cremades}, {Howard}, {Hudson}, {Klassen}, {Klein}, {Lee},
  {Linker}, {Maia}, {Mikic}, {Raymond}, {Reiner}, {Simnett}, {Srivastava},
  {Tripathi}, {Vainio}, {Vourlidas}, {Zhang}, {Zurbuchen}, {Sheeley}, \&
  {Marqu{\'e}}}]{2006cme..book..341P}
{Pick}, M., {Forbes}, T.~G., {Mann}, G., {et~al.} 2006, {Multi-Wavelength
  Observations of CMEs and Associated Phenomena}, Vol.~21 (Kluwer Academic
  Publishers), 341

\bibitem[{{Powell} {et~al.}(1999){Powell}, {Roe}, {Linde}, {Gombosi}, \& {De
  Zeeuw}}]{1999JCoPh.154..284P}
{Powell}, K.~G., {Roe}, P.~L., {Linde}, T.~J., {Gombosi}, T.~I., \& {De Zeeuw},
  D.~L. 1999, Journal of Computational Physics, 154, 284

\bibitem[{{Ramesh} {et~al.}(2012){Ramesh}, {Lakshmi}, {Kathiravan},
  {Gopalswamy}, \& {Umapathy}}]{2012ApJ...752..107R}
{Ramesh}, R., {Lakshmi}, M.~A., {Kathiravan}, C., {Gopalswamy}, N., \&
  {Umapathy}, S. 2012, \apj, 752, 107

\bibitem[{{Reiners}(2014)}]{2014IAUS..302..156R}
{Reiners}, A. 2014, in IAU Symposium, Vol. 302, Magnetic Fields throughout
  Stellar Evolution, ed. P.~{Petit}, M.~{Jardine}, \& H.~C. {Spruit}, 156--163

\bibitem[{{Reiners} \& {Basri}(2008)}]{2008A&A...489L..45R}
{Reiners}, A., \& {Basri}, G. 2008, \aap, 489, L45

\bibitem[{{Sachdeva} {et~al.}(2019){Sachdeva}, {van der Holst}, {Manchester},
  {T{\'o}th}, {Chen}, {Lloveras}, {V{\'a}squez}, {Lamy}, {Wojak}, {Jackson},
  {Yu}, \& {Henney}}]{2019ApJ...887...83S}
{Sachdeva}, N., {van der Holst}, B., {Manchester}, W.~B., {et~al.} 2019, \apj,
  887, 83

\bibitem[{{Schmidt} \& {Cairns}(2016)}]{2016GeoRL..43...50S}
{Schmidt}, J.~M., \& {Cairns}, I.~H. 2016, \grl, 43, 50

\bibitem[{{Segura} {et~al.}(2010){Segura}, {Walkowicz}, {Meadows}, {Kasting},
  \& {Hawley}}]{2010AsBio..10..751S}
{Segura}, A., {Walkowicz}, L.~M., {Meadows}, V., {Kasting}, J., \& {Hawley}, S.
  2010, Astrobiology, 10, 751

\bibitem[{{Sharma} \& {Mittal}(2017)}]{2017Ap.....60..213S}
{Sharma}, J., \& {Mittal}, N. 2017, Astrophysics, 60, 213

\bibitem[{{Shibayama} {et~al.}(2013){Shibayama}, {Maehara}, {Notsu}, {Notsu},
  {Nagao}, {Honda}, {Ishii}, {Nogami}, \& {Shibata}}]{2013ApJS..209....5S}
{Shibayama}, T., {Maehara}, H., {Notsu}, S., {et~al.} 2013, The Astrophysical
  Journal Supplement Series, 209, doi:10.1088/0067-0049/209/1/5

\bibitem[{{Shulyak} {et~al.}(2019){Shulyak}, {Reiners}, {Nagel}, {Tal-Or},
  {Caballero}, {Zechmeister}, {B{\'e}jar}, {Cort{\'e}s-Contreras}, {Martin},
  {Kaminski}, {Ribas}, {Quirrenbach}, {Amado}, {Anglada-Escud{\'e}}, {Bauer},
  {Dreizler}, {Guenther}, {Henning}, {Jeffers}, {K{\"u}rster}, {Lafarga},
  {Montes}, {Morales}, \& {Pedraz}}]{2019A&A...626A..86S}
{Shulyak}, D., {Reiners}, A., {Nagel}, E., {et~al.} 2019, \aap, 626, A86

\bibitem[{{Sindhuja} \& {Gopalswamy}(2020)}]{2020ApJ...889..104S}
{Sindhuja}, G., \& {Gopalswamy}, N. 2020, \apj, 889, 104

\bibitem[{{Sokolov} {et~al.}(2013){Sokolov}, {van der Holst}, {Oran}, {Downs},
  {Roussev}, {Jin}, {Manchester}, {Evans}, \& {Gombosi}}]{2013ApJ...764...23S}
{Sokolov}, I.~V., {van der Holst}, B., {Oran}, R., {et~al.} 2013, \apj, 764, 23

\bibitem[{{Sun} {et~al.}(2015){Sun}, {Bobra}, {Hoeksema}, {Liu}, {Li}, {Shen},
  {Couvidat}, {Norton}, \& {Fisher}}]{2015ApJ...804L..28S}
{Sun}, X., {Bobra}, M.~G., {Hoeksema}, J.~T., {et~al.} 2015, \apjl, 804, L28

\bibitem[{{Thalmann} {et~al.}(2015){Thalmann}, {Su}, {Temmer}, \&
  {Veronig}}]{2015ApJ...801L..23T}
{Thalmann}, J.~K., {Su}, Y., {Temmer}, M., \& {Veronig}, A.~M. 2015, \apjl,
  801, L23

\bibitem[{{Thalmann} {et~al.}(2017){Thalmann}, {Su}, {Temmer}, \&
  {Veronig}}]{2017ApJ...844L..27T}
---. 2017, \apjl, 844, L27

\bibitem[{{Tilley} {et~al.}(2019){Tilley}, {Segura}, {Meadows}, {Hawley}, \&
  {Davenport}}]{2019AsBio..19...64T}
{Tilley}, M.~A., {Segura}, A., {Meadows}, V., {Hawley}, S., \& {Davenport}, J.
  2019, Astrobiology, 19, 64

\bibitem[{{Titov} \& {D{\'e}moulin}(1999)}]{1999A&A...351..707T}
{Titov}, V.~S., \& {D{\'e}moulin}, P. 1999, \aap, 351, 707

\bibitem[{{Toriumi} {et~al.}(2017){Toriumi}, {Schrijver}, {Harra}, {Hudson}, \&
  {Nagashima}}]{2017ApJ...834...56T}
{Toriumi}, S., {Schrijver}, C.~J., {Harra}, L.~K., {Hudson}, H., \&
  {Nagashima}, K. 2017, \apj, 834, 56

\bibitem[{{T{\'o}th} {et~al.}(2012){T{\'o}th}, {van der Holst}, {Sokolov}, {De
  Zeeuw}, {Gombosi}, {Fang}, {Manchester}, {Meng}, {Najib}, {Powell}, {Stout},
  {Glocer}, {Ma}, \& {Opher}}]{2012JCoPh.231..870T}
{T{\'o}th}, G., {van der Holst}, B., {Sokolov}, I.~V., {et~al.} 2012, Journal
  of Computational Physics, 231, 870

\bibitem[{{Tuomi} {et~al.}(2019){Tuomi}, {Jones}, {Anglada-Escud{\'e}},
  {Butler}, {Arriagada}, {Vogt}, {Burt}, {Laughlin}, {Holden}, {Teske},
  {Shectman}, {Crane}, {Thompson}, {Keiser}, {Jenkins}, {Berdi{\~n}as}, {Diaz},
  {Kiraga}, \& {Barnes}}]{2019arXiv190604644T}
{Tuomi}, M., {Jones}, H.~R.~A., {Anglada-Escud{\'e}}, G., {et~al.} 2019, arXiv
  e-prints, arXiv:1906.04644

\bibitem[{{Turner} \& {Drake}(2009)}]{2009ApJ...703.2152T}
{Turner}, N.~J., \& {Drake}, J.~F. 2009, \apj, 703, 2152

\bibitem[{{van der Holst} {et~al.}(2019){van der Holst}, {Manchester}, {Klein},
  \& {Kasper}}]{2019ApJ...872L..18V}
{van der Holst}, B., {Manchester}, W.~B., I., {Klein}, K.~G., \& {Kasper},
  J.~C. 2019, \apjl, 872, L18

\bibitem[{{van der Holst} {et~al.}(2014){van der Holst}, {Sokolov}, {Meng},
  {Jin}, {Manchester}, {T{\'o}th}, \& {Gombosi}}]{2014ApJ...782...81V}
{van der Holst}, B., {Sokolov}, I.~V., {Meng}, X., {et~al.} 2014, \apj, 782, 81

\bibitem[{{van Haarlem} {et~al.}(2013){van Haarlem}, {Wise}, {Gunst}, {Heald},
  {McKean}, {Hessels}, {de Bruyn}, {Nijboer}, {Swinbank}, {Fallows},
  {Brentjens}, {Nelles}, {Beck}, {Falcke}, {Fender}, {H{\"o}randel},
  {Koopmans}, {Mann}, {Miley}, {R{\"o}ttgering}, {Stappers}, {Wijers},
  {Zaroubi}, {van den Akker}, {Alexov}, {Anderson}, {Anderson}, {van Ardenne},
  {Arts}, {Asgekar}, {Avruch}, {Batejat}, {B{\"a}hren}, {Bell}, {Bell}, {van
  Bemmel}, {Bennema}, {Bentum}, {Bernardi}, {Best}, {B{\^\i}rzan}, {Bonafede},
  {Boonstra}, {Braun}, {Bregman}, {Breitling}, {van de Brink}, {Broderick},
  {Broekema}, {Brouw}, {Br{\"u}ggen}, {Butcher}, {van Cappellen}, {Ciardi},
  {Coenen}, {Conway}, {Coolen}, {Corstanje}, {Damstra}, {Davies}, {Deller},
  {Dettmar}, {van Diepen}, {Dijkstra}, {Donker}, {Doorduin}, {Dromer}, {Drost},
  {van Duin}, {Eisl{\"o}ffel}, {van Enst}, {Ferrari}, {Frieswijk}, {Gankema},
  {Garrett}, {de Gasperin}, {Gerbers}, {de Geus}, {Grie{\ss}meier}, {Grit},
  {Gruppen}, {Hamaker}, {Hassall}, {Hoeft}, {Holties}, {Horneffer}, {van der
  Horst}, {van Houwelingen}, {Huijgen}, {Iacobelli}, {Intema}, {Jackson},
  {Jelic}, {de Jong}, {Juette}, {Kant}, {Karastergiou}, {Koers}, {Kollen},
  {Kondratiev}, {Kooistra}, {Koopman}, {Koster}, {Kuniyoshi}, {Kramer},
  {Kuper}, {Lambropoulos}, {Law}, {van Leeuwen}, {Lemaitre}, {Loose}, {Maat},
  {Macario}, {Markoff}, {Masters}, {McFadden}, {McKay-Bukowski}, {Meijering},
  {Meulman}, {Mevius}, {Middelberg}, {Millenaar}, {Miller-Jones}, {Mohan},
  {Mol}, {Morawietz}, {Morganti}, {Mulcahy}, {Mulder}, {Munk}, {Nieuwenhuis},
  {van Nieuwpoort}, {Noordam}, {Norden}, {Noutsos}, {Offringa}, {Olofsson},
  {Omar}, {Orr{\'u}}, {Overeem}, {Paas}, {Pand ey-Pommier}, {Pandey}, {Pizzo},
  {Polatidis}, {Rafferty}, {Rawlings}, {Reich}, {de Reijer}, {Reitsma},
  {Renting}, {Riemers}, {Rol}, {Romein}, {Roosjen}, {Ruiter}, {Scaife}, {van
  der Schaaf}, {Scheers}, {Schellart}, {Schoenmakers}, {Schoonderbeek},
  {Serylak}, {Shulevski}, {Sluman}, {Smirnov}, {Sobey}, {Spreeuw}, {Steinmetz},
  {Sterks}, {Stiepel}, {Stuurwold}, {Tagger}, {Tang}, {Tasse}, {Thomas},
  {Thoudam}, {Toribio}, {van der Tol}, {Usov}, {van Veelen}, {van der Veen},
  {ter Veen}, {Verbiest}, {Vermeulen}, {Vermaas}, {Vocks}, {Vogt}, {de Vos},
  {van der Wal}, {van Weeren}, {Weggemans}, {Weltevrede}, {White}, {Wijnholds},
  {Wilhelmsson}, {Wucknitz}, {Yatawatta}, {Zarka}, {Zensus}, \& {van
  Zwieten}}]{2013A&A...556A...2V}
{van Haarlem}, M.~P., {Wise}, M.~W., {Gunst}, A.~W., {et~al.} 2013, \aap, 556,
  A2

\bibitem[{{Vida} {et~al.}(2019){Vida}, {Leitzinger}, {Kriskovics}, {Seli},
  {Odert}, {Kov{\'a}cs}, {Korhonen}, \& {van
  Driel-Gesztelyi}}]{2019A&A...623A..49V}
{Vida}, K., {Leitzinger}, M., {Kriskovics}, L., {et~al.} 2019, \aap, 623, A49

\bibitem[{{Vidotto} {et~al.}(2014){Vidotto}, {Jardine}, {Morin}, {Donati},
  {Opher}, \& {Gombosi}}]{2014MNRAS.438.1162V}
{Vidotto}, A.~A., {Jardine}, M., {Morin}, J., {et~al.} 2014, \mnras, 438, 1162

\bibitem[{{Villadsen} \& {Hallinan}(2019)}]{2019ApJ...871..214V}
{Villadsen}, J., \& {Hallinan}, G. 2019, \apj, 871, 214

\bibitem[{{Villadsen}(2017)}]{2017PhDT.........8V}
{Villadsen}, J.~R. 2017, PhD thesis, California Institute of Technology

\bibitem[{{Wargelin} \& {Drake}(2002)}]{2002ApJ...578..503W}
{Wargelin}, B.~J., \& {Drake}, J.~J. 2002, \apj, 578, 503

\bibitem[{{Webb} \& {Howard}(2012)}]{2012LRSP....9....3W}
{Webb}, D.~F., \& {Howard}, T.~A. 2012, Living Reviews in Solar Physics, 9,
  doi:10.12942/lrsp-2012-3

\bibitem[{{Wild} \& {McCready}(1950)}]{1950AuSRA...3..387W}
{Wild}, J.~P., \& {McCready}, L.~L. 1950, Australian Journal of Scientific
  Research A Physical Sciences, 3, 387

\bibitem[{{Wood}(2018)}]{2018JPhCS1100a2028W}
{Wood}, B.~E. 2018, in Journal of Physics Conference Series, Vol. 1100, Journal
  of Physics Conference Series, 012028

\bibitem[{{Wood} {et~al.}(2001){Wood}, {Linsky}, {M{\"u}ller}, \&
  {Zank}}]{2001ApJ...547L..49W}
{Wood}, B.~E., {Linsky}, J.~L., {M{\"u}ller}, H.-R., \& {Zank}, G.~P. 2001,
  \apjl, 547, L49

\bibitem[{{Wood} {et~al.}(2005){Wood}, {M{\"u}ller}, {Zank}, {Linsky}, \&
  {Redfield}}]{2005ApJ...628L.143W}
{Wood}, B.~E., {M{\"u}ller}, H.-R., {Zank}, G.~P., {Linsky}, J.~L., \&
  {Redfield}, S. 2005, \apjl, 628, L143

\bibitem[{{Yadav} {et~al.}(2016){Yadav}, {Christensen}, {Wolk}, \&
  {Poppenhaeger}}]{2016ApJ...833L..28Y}
{Yadav}, R.~K., {Christensen}, U.~R., {Wolk}, S.~J., \& {Poppenhaeger}, K.
  2016, \apjl, 833, L28

\bibitem[{{Yashiro} \& {Gopalswamy}(2009)}]{2009IAUS..257..233Y}
{Yashiro}, S., \& {Gopalswamy}, N. 2009, in IAU Symposium, Vol. 257, Universal
  Heliophysical Processes, ed. N.~{Gopalswamy} \& D.~F. {Webb}, 233--243

\bibitem[{{Yi{\v{g}}it}(2018)}]{2018assi.book.....Y}
{Yi{\v{g}}it}, E. 2018, {Atmospheric and Space Sciences: Ionospheres and Plasma
  Environments}, Vol.~2, doi:10.1007/978-3-319-62006-0

\bibitem[{{Zucca} {et~al.}(2014){Zucca}, {Carley}, {Bloomfield}, \&
  {Gallagher}}]{2014A&A...564A..47Z}
{Zucca}, P., {Carley}, E.~P., {Bloomfield}, D.~S., \& {Gallagher}, P.~T. 2014,
  \aap, 564, A47

\end{thebibliography}



\end{document}